\documentclass[DIV=calc, paper=a4, fontsize=11pt, onecolumn]{scrartcl}
\usepackage{amsfonts,amssymb,amsmath}
 \usepackage{lineno,hyperref}
\usepackage{graphicx}
\usepackage{subfig}
\usepackage{float}
\usepackage{tabularx}
\usepackage{authblk} 
\usepackage{arydshln,leftidx,mathtools}
\usepackage{accents}
\usepackage[noend]{algpseudocode}
\modulolinenumbers[1]

\usepackage{xcolor}   

\usepackage[english]{babel}
\usepackage[utf8]{inputenc}
\usepackage{parskip}
\usepackage{multirow, float}
\usepackage[top=2.5cm, left=3cm, right=3cm, bottom=4.0cm]{geometry}

 \title{
A weather-driven mathematical model of \textit{Culex} population abundance and the impact of vector control interventions
  }
\author[1]{Suman Bhowmick\footnote{Corresponding Author}}
\author[4, 5]{Patrick Irwin}
\author[7]{Kristina Lopez}
\author[6]{Megan Lindsay Fritz}
\author[1, 2, 3]{Rebecca Lee Smith}
\affil[1]{Department of Pathobiology, University of Illinois at Urbana-Champaign, Urbana, Illinois, USA}
\affil[2]{Carl R. Woese Institute for Genomic Biology, University of Illinois at Urbana-Champaign, Urbana, Illinois, USA}
\affil[3]{Carle Illinois College of Medicine, University of Illinois at Urbana-Champaign, Urbana, Illinois, USA}
\affil[4]{Department of Entomology, University of Wisconsin-Madison, Madison, Wisconsin, USA}
\affil[5]{Department of Pathobiological Sciences, University of Wisconsin-Madison, Madison, Wisconsin, USA}
\affil[6]{Department of Entomology, Institute for Advanced Computer Studies, University of Maryland, USA}
\affil[7]{North Shore Mosquito Abatement District, Northfield, Illinois, USA}

\begin{document}
\maketitle
\tableofcontents
\section{Highlights}
\begin{itemize}
\item We build and calibrate a mathematical model for the dynamics of mosquito population by incorporating entomological data and weather-driven factors.
\item We fit the weather-driven Ordinary Differential Equation model to local mosquito trap data. 
\item We perform sensitivity analysis of \textit{Basic offspring number}: $R_{0_{F}}$.
\item We assess the effectiveness of the vector control approach in Cook County.
\end{itemize}
\section{Abstract}
Even as the incidence of mosquito-borne diseases like West Nile Virus (WNV) in North America has risen over the past decade, effectively modelling mosquito population density or, the abundance has proven to be a persistent challenge.
It is critical to capture the fluctuations in mosquito abundance across seasons in order to forecast the varying risk of disease transmission from one year to the next.
We develop a process-based mechanistic weather-driven Ordinary Differential Equation (ODE) model to study the population biology of both aqueous and terrestrial stages of mosquito population. 
The progression of mosquito lifecycle through these stages is influenced by different factors, including temperature, daylight hours, intra-species competition and the availability of aquatic habitats.
Weather-driven parameters are utilised in our work, are a combination of laboratory research and literature data.  
In our model, we include precipitation data as a substitute for evaluating additional mortality in the mosquito population.
We compute the \textit{Basic offspring number} of the associated model and perform sensitivity analysis.
Finally, we employ our model to assess the effectiveness of various adulticides strategies to predict the reduction in mosquito population.
This enhancement in modelling of mosquito abundance can be instrumental in guiding interventions aimed at reducing mosquito populations and mitigating mosquito-borne diseases such as the WNV.

\section{Introduction}\label{Introduction}
Mosquito-borne diseases pose significant threats to public health globally, necessitating the development of effective strategies for vector control \cite{FRANKLINOS2019e302}.
The dynamics of mosquito populations directly impact the spread of diseases transmitted by mosquitoes, including West Nile virus (WNV), zika, malaria, dengue, and chikungunya \cite{201827, Joachim, Colon, Erazo}.
Climate change is undeniably among the paramount challenges influencing our world at present, and the altering climate, global warming  notably worsens the threat of mosquito-borne disease transmission \cite{Joachim, doi:10.1098/rstb.2013.0552}.
Each year, vector-borne diseases result in hundreds of millions of cases worldwide and diseases like malaria, dengue, and WNV represent over $17\%$ of all infectious diseases and are responsible for more than $700,000$ deaths each year \cite{WHO, NIH}.
In the age of climate change and global warming, gaining insight into the impact of climate variables (temperature, precipitation, humidity, and wind etc. ) on mosquito population dynamics is becoming crucial \cite{doi:10.1289/ehp.01109s1141, deSouza, BRUGUERAS2020110038, doi:10.1056/NEJMra2200092, EPA1}. 
This understanding is essential to inform public health policies aimed at controlling diseases \cite{Braith, 10.1093/jme/tjab219, connelly2020continuation, doi:10.2105/AJPH.2008.156224}.

Mosquitoes belonging to the \textit{Culex} genus, within the Culicidae family, serve as significant carriers for different infectious diseases, such as Rift Valley fever, St. Louis encephalitis, Japanese encephalitis, and West Nile fever \cite{Moser, 10.1603/ME13003, Weaver}.
Functioning as vectors, they play an important role in the transmission cycles of viruses, thereby contributing to the spread of various viral infections \cite{HONGOH201253, 10.1093/jme/tjz151}. 
Therefore, to comprehend the eco-epidemiology of mosquito-borne arboviruses, it is crucial to have an understanding of the population dynamics of these vectors \cite{Allan, https://doi.org/10.1029/2022GH000708, 10.7554/eLife.58511, CAILLY20127}.
In North America, the primary vectors for WNV are \textit{Culex} mosquitoes, particularly \textit{Cx. pipiens}, \textit{Cx. tarsalis}, \textit{Cx. quinquefasciatus}, and \textit{Cx. restuans}. 
Given their geographical distribution, \textit{Cx. pipiens} serves as the predominant vector in the northern regions, \textit{Cx. tarsalis} in the western regions, and \textit{Cx. quinquefasciatus} in the southern regions of North America \cite{10.1093/jme/tjz146, Gorris1}.

As exothermic organisms, mosquitoes are exceptionally responsive to fluctuations in ambient temperature. 
The influence of temperature on various characteristics of mosquito biology, including developmental rate, mortality, life-history traits, diapause, and oviposition, has been extensively investigated and continues to be a subject of intense research \cite{10.1603/ME13003, clements2023biology, 10.7554/eLife.58511}.
Climatic factors, such as air temperature and precipitation, have a direct and non-linear effect on mosquito populations \cite{VALDEZ201728, Federico, 10.1603/ME10117, 10.1603/ME11073}. 
Mosquito survival and development require a sufficiently warm ambient temperature, but excessively high temperatures can yield to increased mortality \cite{https://doi.org/10.1111/j.1749-6632.2001.tb02699.x, 10.1603/ME13003, Moser}. 
Adequate precipitation is an essential factor for creating suitable egg-laying sites for juvenile mosquitoes; however, excessive rainfall poses the risk of flushing eggs and larvae from their habitats \cite{VALDEZ201728, SOH2021142420}.
Studies conducted in laboratory settings have demonstrated the importance of temperature for the development time and survival rates of eggs, larvae, and pupae \cite{10.1093/jmedent/45.1.28, SOH2021142420}.
Temperature has an immense impact on the lifespan, duration of the gonotrophic cycle, and the rate of virus transmission in adult mosquitoes \cite{Moser, 10.7554/eLife.58511}.
Certain \textit{Culex} mosquitoes experience reproductive diapause, a phenomenon influenced by both temperature and duration of daylight \cite{annurev:/content/journals/10.1146/annurev-ento-011613-162023, LAPERRIERE201199}.
Therefore, incorporating multiple weather-driven factors is essential for developing a robust and comprehensive model.\par

Understanding the intricate relationship between meteorological variables and mosquito population dynamics is crucial for devising targeted and efficient vector control interventions \cite{10.1093/jme/tjz083}.
Mathematical modelling is crucial for developing effective mosquito abatement strategies by simulating mosquito population dynamics, host-vector disease transmission patterns \cite{Demers}.
These models incorporate different weather-driven factors and help to identify optimal times for interventions like applying larvicide and adulticide \cite{BHOWMICK2024107346}.
Additionally, predictive modelling can potentially help in early warning systems and scenario analysis, thus ensuring proactive and efficient responses to potential disease outbreaks \cite{10.1093/jme/tjz083, BHOWMICK2024107346}.
Our study explores the impact of various vector control intervention strategies on the modelled mosquito abundance. 
We evaluate the efficacy of different strategies of spraying of adulticide in mitigating mosquito-borne disease transmission, practiced by the North West Mosquito Abatement District (NWMAD), Chicago, Cook County, Illinois, USA \cite{10.1093/jme/tjad088, 10.2987/19-6848.1}. 
Through a systematic way of simulations, we aim to provide insights into the optimal timing, frequency, and combination of interventions to achieve maximum effectiveness.

The development of predictive models and the assessment of control measures become paramount as we can witness that the global community faces ongoing challenges from emerging infectious diseases transmitted by mosquitoes.
Various mathematical models have been developed to study the population dynamics of \textit{Culex} mosquitoes.
These models differ in complexity and scope, encompassing simple compartmental models, more advanced network-based metapopulation models, as well as statistical and machine learning-based approaches \cite{10.1093/jme/tjac127, Gorris, Gorris1, 10.3389/fvets.2024.1383320, Lebl, Kollas, Clare, BHOWMICK2023110213}.
Compartment-based models are especially useful for modeling age-structured populations, as they divide the population into distinct groups based on life stages, making them straightforward to implement \cite{FRANTZ2024110764}. 
This segmentation enables the analytical calculation of the Basic Offspring Number ($R_{0_{F}}$) and allows for simulations to evaluate the effectiveness of various control strategies, as well as predict the progression of the mosquito life cycle.
We explore \textit{Culex} mosquito population dynamics and evaluate various vector control interventions extensively through a weather-driven mathematical modelling approach.
Here, we include key contributions in the field that have laid the groundwork for our weather-driven mathematical model.
Early deterministic models by \cite{EZANNO201539, CAILLY20127, ijerph10051698} incorporated temperature-dependent life cycle parameters to simulate mosquito population dynamics. 
These models highlight the critical role of temperature in influencing different traits of \textit{Culex} population and explore two different abatement strategies in a theoretical framework.
The authors in \cite{Ewing} develop a deterministic model that accounted for seasonal variations in rainfall, demonstrating how density-independent mortality and interspecific predation interact  drive the population cycles of \textit{Culex} mosquitoes.
The authors in \cite{YU201828} devise a compartmental model to estimate the abundance of \textit{Culex} and validated their results against trap data but the model was, however, solely temperature-driven. 
It is to be noted that other weather-driven factors such as precipitation, landscape, and wind can potentially improve the predictive nature of future models.
The authors in \cite{insects14030293} propose a compartmental based model that considers both precipitation and temperature.
However, the authors do not account for the impact of flushing in catch basins following heavy rainfall \cite{10.1093/jmedent/45.1.28, SOH2021142420}. 
The aforementioned studies notably overlook the incorporation of one important epidemiological metric Basic Offspring number and its temperature dependence. 
Basic Offspring number  constitutes of several temperature-driven traits that play an important role in shaping the \textit{Culex} population dynamics as mosquitoes are exothermic organisms \cite{10.7554/eLife.58511}. 
The studies mentioned above also do not examine the impact of various abatement strategies employed in different mosquito abatement districts, nor do they compare these strategies with the relative abundance of mosquitoes based on the pooled data and demonstrating them visually.
Therefore, a more comprehensive and realistic approach is needed to address the impact of adulticide in mosquito abatement strategies in the presence of weather-driven mosquito parameters.

\par
In our current work, we present a comprehensive investigation into the dynamics of mosquito population abundance, governed by a weather-driven process based mathematical model.
The influence of weather-driven factors on mosquito life cycles has been well-established, with temperature and precipitation playing pivotal roles in modulating breeding, development, and survival rates.
By harnessing the power of mathematical modelling, this research endeavour to develop a comprehensive and weather-driven mathematical model to better understand the dynamics of \textit{Culex} populations and our model offers a dynamic and adaptive framework for predicting mosquito population abundance.


\section{Data collection}\label{Datacollection}

\subsection{Mosquito trap data}\label{Mosquitotrapdata}
\begin{figure}[H]
\centering
\includegraphics[width=12cm]{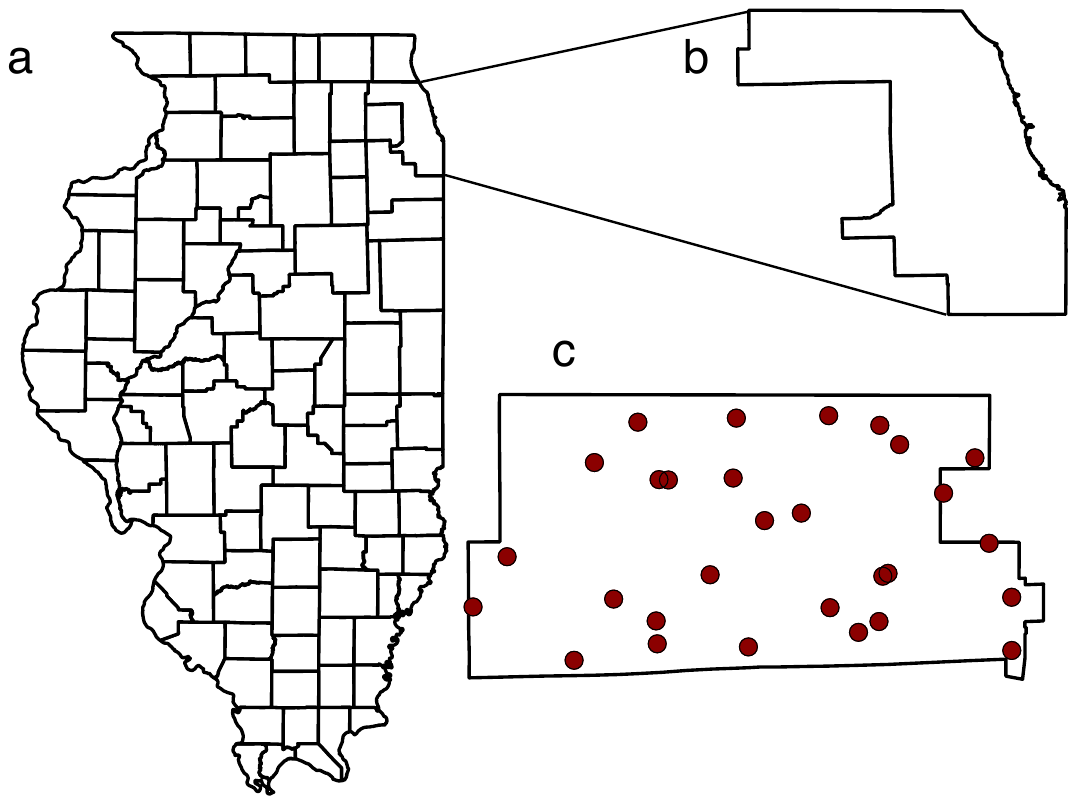}
	\caption{ \textit{Culex} population model \eqref{Eq1} study area, displaying the locations of mosquito trap data.
	Figure (c) shows the locations of the trap data, represented by the red dots under the jurisdiction of NWMAD within the Cook county (Figure (b)), Illinois, USA (Figure (a))
	 serviced between 2014 and 2019.
           }
    \label{fig:TrapData}
\end{figure}

Data on mosquito testing from 2014 to 2019 are acquired from the Illinois Department of Public Health (IDPH) through a user agreement \cite{IDPH}. 
The IDPH consolidates information gathered by local public health agencies and mosquito abatement districts throughout Illinois, managing a comprehensive statewide database for mosquito testing results.
The IDPH has set up a mosquito surveillance protocol for local health and mosquito abatement districts to follow, while ensuring standardised mosquito collection and testing statewide. 
Commonly, the local agencies employ gravid traps to collect vector mosquitoes, determine the mosquitoes' sex and species, and create pools comprising up to 50 mosquitoes of a specific species from those captured in each trap. 
If fewer than 50 mosquitoes are collected, the pool will contain the actual number captured.
Our analysis explicitly utilise test results from pools of female \textit{Culex} mosquitoes. 
Not all mosquitoes undergo species identification before testing; nonetheless, it is noteworthy that the predominant species among the \textit{Culex} mosquitoes collected in this region are \textit{Cx. pipiens} or \textit{Cx. restuans}.
In our study, we assume that climate and weather changes have comparable effects on the dynamics of these species \cite{10.1603/ME13003}.
The data from the mosquito trap was gathered at the North West Mosquito Abatement District (NWMAD), an area spanning 605 square kilometres, encompassing the northwest suburbs of Chicago in Cook County, Illinois, USA (see Figure \ref{fig:TrapData}).

\subsection{Weather data}\label{Weatherdata}

\begin{figure}[H]
\centering
\subfloat[Mean daily temperature data]{\includegraphics[width=0.45\textwidth]{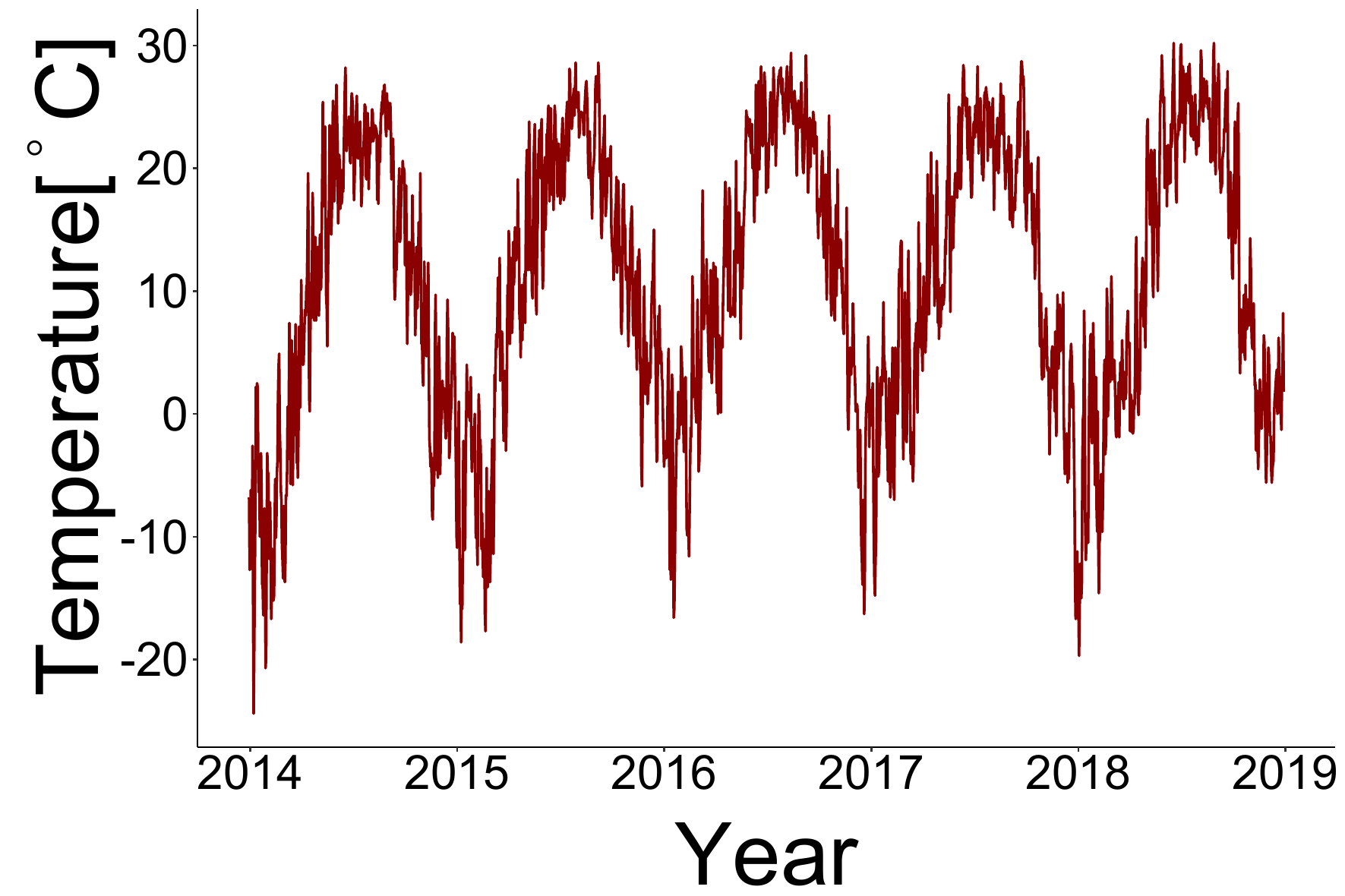}\label{fig:f1}}
\hfill
\subfloat[Precipitation data]{\includegraphics[width=0.45\textwidth]{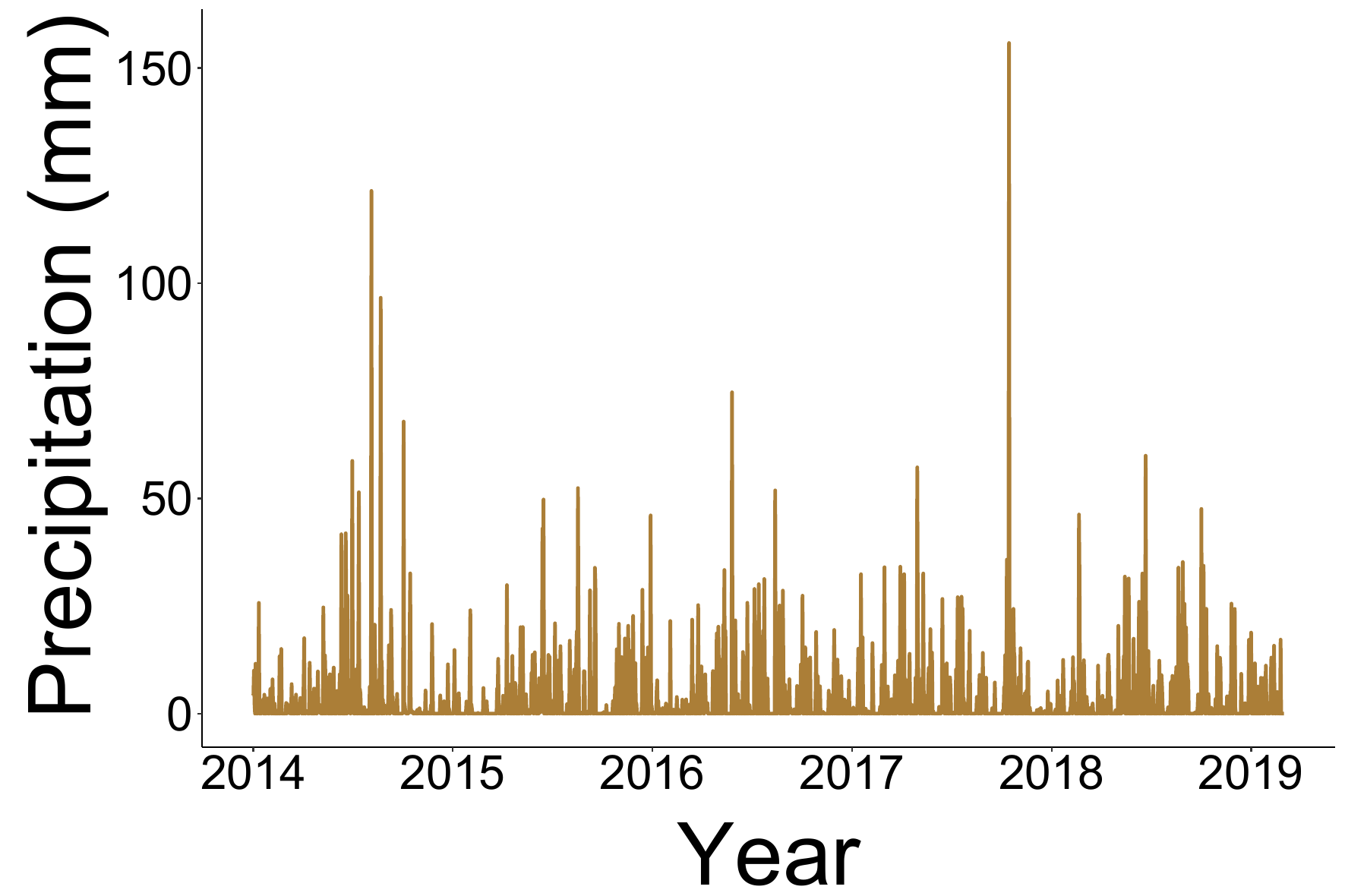}\label{fig:f2}}
\caption{Time series of daily mean temperature and precipitation data for the NWMAD county spanning from 2014 to 2019 provided by the PRISM Climate Group, affiliated with Oregon State University \cite{PRISMClimateGroup}.
}\label{fig:WeatherData}
\end{figure}

We collect daily mean temperature and precipitation data for the period 2014 to 2019 from the PRISM Climate Group \cite{PRISMClimateGroup},  and the Figure \ref{fig:WeatherData} illustrates the time-series of temperature-precipitation data during this period. 
The PRISM daily temperature data is available in 4 km resolution spatial grids, computed using interpolation and statistical methods. 
These methods enable the integration of point data from weather monitoring networks across the nation with topographic data.

\section{Model description}\label{Modeldescription} 
\subsection{\textit{Culex} lifecycle}\label{Lifecycle}
The life cycle of \textit{Culex} mosquitoes encompasses aquatic stages (egg, larva, and pupa) and terrestrial stages (adult), as depicted in the Figure \ref{fig:ModelFlow} \cite{https://doi.org/10.1111/j.1749-6632.2001.tb02699.x, EPA}.
Following the final aquatic stage, male and female adults undergo swift mating upon emergence.
After insemination, female \textit{Culex} mosquitoes disperse to locate a blood feeding host, potentially engaging in long-distance movements with the associated risk of encountering host defence responses \cite{EZANNO201539}. 
After obtaining a blood meal, females typically seek refuge in a sheltered location for the few days required for egg maturation \cite{EZANNO201539, EPA}. 
\textit{Culex} mosquitoes exhibit movement patterns involving ingress and egress from the resting sites, thus resulting in localised and less risky movements. 
Subsequently, female  \textit{Culex} mosquitoes search for an oviposition site, which can possibly involve another round of long-distance and risky movements. 
Depending on the mosquito species, various sites may be utilised, ranging from aquatic environments to humid places. 
Hatching can occur shortly after laying eggs or may be delayed for several months, depending on the species and the time of year when the egg are deposited. 
The larvae then progress through four larval stages before transitioning into pupae, eventually emerging as adults on the water's surface \cite{EPA, EZANNO201539, CAILLY20127}.

\subsection{Mathematical model}\label{Mathematicalmodel}
The mathematical model we constructed for \textit{Culex} population dynamics possesses several characteristics:
(i) Mechanistic: It includes a priori mathematical descriptions for all mosquito-related processes in a simplified approach.
(ii) Deterministic: Our model depicts the average behaviour of the population, making it well-suited for large populations, such as those formed by mosquitoes.
(iii) Weather-driven: Given that mosquitoes are poikilotherms and cannot regulate their body temperature, our model is developed to be highly dependent on different weather-driven conditions \cite{CAILLY20127, EZANNO201539, ijerph10051698,  YU201828}.
(iv) Overwintering consideration: Our model explicitly takes into account of overwintering processes, thus recognising their significance in the overall dynamics and this makes the model more relevant \cite{BHOWMICK2020110117, LAPERRIERE201199}. 

\begin{figure}[H]
\centering
\includegraphics[width=12cm]{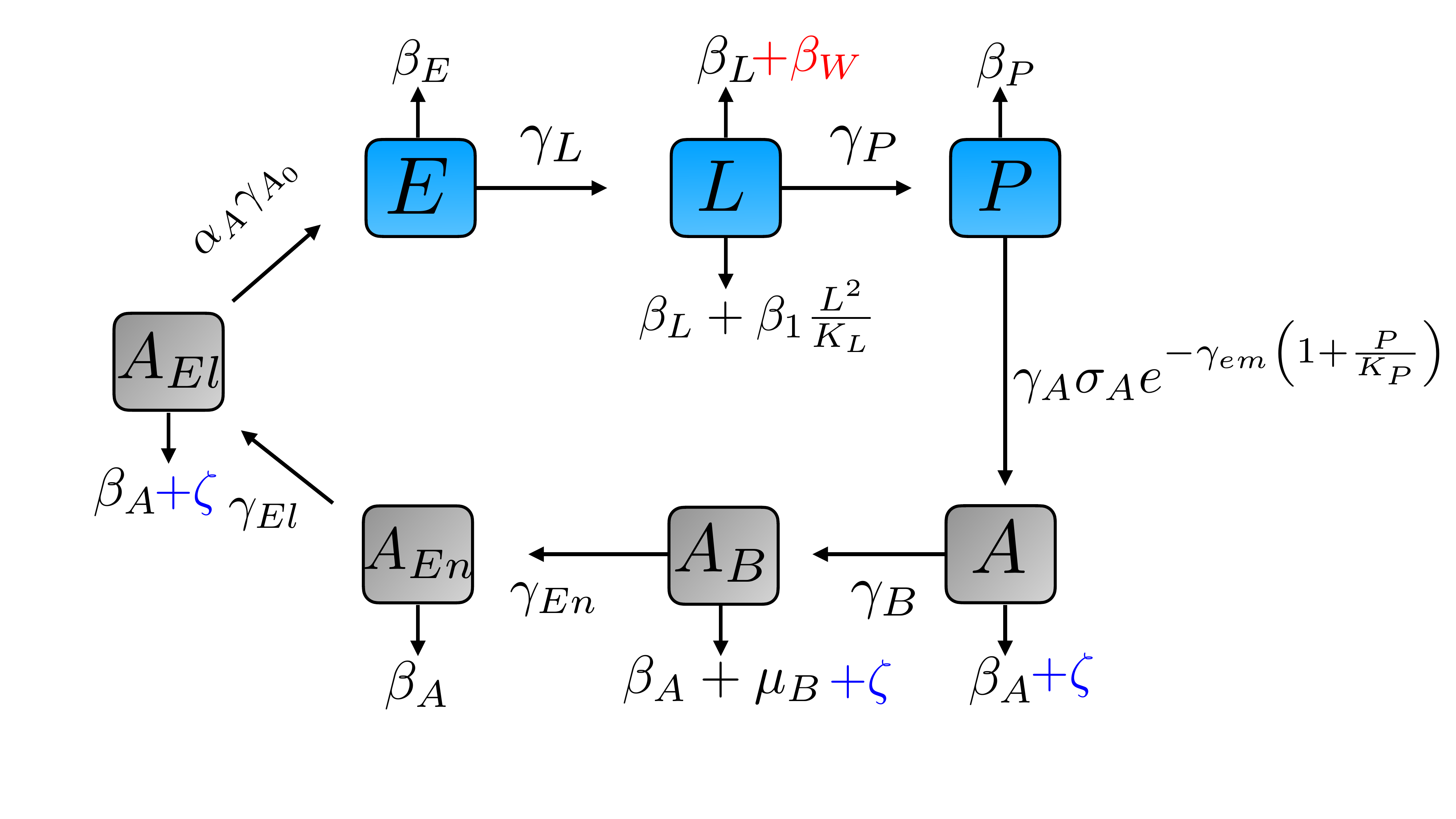}
	\caption{ Model flow as described in \eqref{Eq1}
Model diagram of \textit{Culex} population dynamics in temperate climate. 
Aquatic stages are drawn in blue, adult females in grey. 
We depict an additional mortality rate due to flushing in larvae population in red ($\beta_W$) and an extra death due to the application of adulticide in blue ($\zeta$) as described in the section \ref{Mathematicalmodel} and \ref{ULVSpray1}.
           }
    \label{fig:ModelFlow}
\end{figure}

We constructed the generic model of \textit{Culex} population by following the approach outlined by the authors in \cite{CAILLY20127, EZANNO201539, YU201828}.
Our generic simplified mathematical model for the \textit{Culex} population provides a comprehensive representation of all stages in the mosquito life cycle (refer to Figure \ref{fig:ModelFlow}).
The model encompasses seven distinct stages for \textit{Culex}, comprising three aquatic stages ($E$ for eggs, $L$ for larvae, $P$ for pupae), one emerging adult stage ($A$), one blood-seeking stage ($A_B$), one engorged stage ($A_{En}$), and one egg-laying stage ($A_{El}$). 
It's important to note that the adult stage representation is specifically focused on females.
After going through parous phase, female \textit{Culex} undergo successive gonotrophic cycles until their death.
Transitions between individual stages in our simplified mathematical model (Figure \ref{fig:ModelFlow}) result from various events that are also weather-driven, including egg mortality or hatching, larval mortality, pupation (the moult of larvae to pupae), pupa mortality, adult emergence, mortality, engorgement, egg maturation, and oviposition \cite{CAILLY20127, EZANNO201539, YU201828, BHOWMICK2020110117, LAPERRIERE201199, ijerph10051698}.
Additionally, our model incorporates density-dependent mortality at the larval stage, a phenomenon commonly observed \cite{clements2023biology, 10.1603/0022-2585-37.5.732, 10.1603/ME13159}.
We take into account the success of adult  \textit{Culex} emergence, which is contingent on pupa density after following \cite{CAILLY20127, EZANNO201539, YU201828}.
We do not consider adult male \textit{Culex} mosquito explicitly into our model \cite{CAILLY20127, EZANNO201539, YU201828}.

During winter, \textit{Culex} species employs a hibernal dormancy known as diapause to withstand unfavourable weather conditions \cite{10.1603/ME13159, annurev:/content/journals/10.1146/annurev-ento-011613-162023, clements2023biology}.
The primary trigger for initiating diapause is the photoperiod, which refers to the duration of light exposure within a 24-hour period. 
When the photoperiod drops below a certain threshold value, a fraction of developing mosquitoes undergo physiological and behavioural alterations that enable them to survive through winter.
While the photoperiod regulates diapause induction, temperature amplifies this response, thus leading to a higher incidence of diapause in mosquitoes as temperatures decrease \cite{annurev:/content/journals/10.1146/annurev-ento-011613-162023}. 
However, precise information on this interaction are currently unbeknownst to us.
\textit{Culex} species enter diapause as adults \cite{clements2023biology}. 
Diapause-bound females begin seeking shelter shortly after emerging and mating, before feeding on blood \cite{clements2023biology, YU201828, annurev:/content/journals/10.1146/annurev-ento-011613-162023}. 
For simplicity in modelling, we designate a specific period during which all individuals in the diapause stage are assumed to overwinter \cite{BHOWMICK2020110117, LAPERRIERE201199}.
This favourable period is defined by boolean variable ($ \psi$) that governs the start and end dates \cite{CAILLY20127, EZANNO201539, YU201828, annurev:/content/journals/10.1146/annurev-ento-011613-162023}.

\begin{eqnarray}\label{Eq1}
 \frac{dE}{dt} &=& \gamma_{A_{0}}\alpha_A A_{El} -  \psi\gamma_L E  -  {\beta_E E}\\ \nonumber
\frac{dL}{dt} &=& \psi\gamma_L E- \beta_L L- \beta_1 \frac{L^2}{K_L}- \gamma_PL-\beta_W L\\ \nonumber
\frac{dP}{dt}&=& \gamma_{P}L-\beta_P P -\gamma_A P\\ \nonumber
\frac{dA}{dt} &=& \gamma_A\sigma_A Pe^{-\gamma_{em}(1+\frac{P}{K_P})}-\beta_A A-\gamma_{B}A\\ \nonumber
\frac{dA_B}{dt}&=& \gamma_{B}A-\beta_A A_B-\mu_B A_B-\gamma_{En}A_B\\ \nonumber
\frac{dA_{En}}{dt}&=& \gamma_{En}A_B-\beta_A A_{En}-\gamma_{El}A_{En}\\\nonumber
\frac{dA_{El}}{dt} &=&\gamma_{El}A_{En}-\beta_A A_{El}
\end{eqnarray}

with 
\[
    \psi= 
\begin{cases}
    0,& \text{During diapause} \\
    1,& \text{otherwise}
\end{cases}
\]

In the Cook county, Illinois, mosquitoes spend a significant amount of time in a year in a state of diapause due to a long period of cold weather from fall to spring. 
The success of diapausing mechanism has an immense effect on the size of the mosquito population the following season and hence on the potential outbreak of WNV \cite{annurev:/content/journals/10.1146/annurev-ento-011613-162023, YU201828}.
There are several factors that affect the survival of diapausing \textit{Culex} mosquitoes such as temperature, precipitation, habitat type etc \cite{annurev:/content/journals/10.1146/annurev-ento-011613-162023}. 
The availability of limited or no information about the number overwintering mosquitoes compels us to take certain assumptions that we describe in the section \ref{Modelvalidation} during our simulation.
When more data will be available, our model is capable to incorporate such further information.
Following the authors in \cite{CAILLY20127, EZANNO201539}, we also include the environmental carrying capacity that limits the \textit{Culex} population growth through a density-dependent manner, fraction of serving emerging pupae transfer to the emerging adults stage.
We follow the authors in to include the density-dependent survival rate of pupae stage and an additional mortality rate associated with the risky host seeking behaviour \cite{CAILLY20127, EZANNO201539}.
Lastly, we also include another additional mortality rate due to flushing when there is a total 2.5 cm rain over a 7 day period in larvae \cite{10.1603/ME11073, 10.1371/journal.pntd.0006935, doi:10.4137/EHI.S24311}.





\begin{table}[H]
\centering
\begin{tabular}{||c c  ||} 
 \hline
 Variables & Definition  \\ [0.5ex] 
 \hline\hline
$E$ & Egg density   \\ 
$L$ & Larvae   density\\
$P$ & Pupae density \\
$A$ & Emerging adult mosquito density   \\ 
$A_B$ & Blood seeking adult mosquito density   \\
$A_{En}$ & Engorged adult mosquito density   \\ 
$A_{El}$ & Egg laying adult mosquito density   \\  
[1ex] 
 \hline
\end{tabular}
\caption{Model variables and their definitions as mentioned in \eqref{Eq1}.}
\label{table:A}
\end{table}

\begin{table}[H]
\centering
\begin{tabular}{||c c c||} 
 \hline
 Parameters & Definition   & Values \\ [0.5ex] 
 \hline\hline
$\alpha_A$    		& Number of eggs laid by egg laying mosquitoes		 		& $350$\\ 
$\beta_E$     		& Natural mortality rate of egg 						 		& $0.0262$\\
$\gamma_L$ 		& Developmental rate of eggs into larvae 				 		& $f(T)$\\
$\gamma_P$ 		& Developmental rate of larvae into pupae 					& $f(T)$\\
$\beta_L$      		& Minimum mortality rate of larvae 							& $0.0304$\\
$\beta_1$     		& Density dependent mortality rate of larvae 					& $f(T)$\\
$\beta_W$     		& Mortality rate of larvae due to flushing  						& $0.003$\\
$\kappa_L$       	& Standard environment carrying capacity for larvae		 		& $8\times10^{8}$\\
$\gamma_A$ 		& Developmental rate of pupae into adult mosquito 				& $f(T)$\\
$\beta_P$     		& Minimum mortality rate of pupae 							& $0.0146$\\
$\kappa_P$      		& Standard environment carrying capacity for pupae 				& $1\times10^{7}$\\
$\beta_A$     		& Natural mortality rate of adult mosquito 						& $f(T)$\\
$\gamma_B$ 		& Developmental rate of adult mosquito into blood seeking 		& $1.143$\\
$\mu_B$      		& Mortality rate of adult mosquito related to seeking behaviour 		& $0.8$ \\ 
$\gamma_{En}$ 	& Developmental rate of blood seeking mosquito into engorged 	& $2$\\
$\gamma_{El}$  	& Developmental rate of engorged mosquito into egg laying  		& $2$\\
$\sigma_A$      		& Sex-ratio at the emergence 								& $0.1$\\
$\gamma_{em}$       & Mortality rate during adult emergence 						& $0.1$\\
$\gamma_{A_{0}}$ 	& Egg laying rate 										& $f(T)$\\
$\zeta_0$ 			& Mortality rate of mosquito due to adulticide 					& $0.5$\\
[1ex] 
 \hline
\end{tabular}
\caption{Parameters of the model \eqref{Eq1}, \eqref{Eq1Modified} and their corresponding values. 
In this context, $f(T)$ represents the function describing temperature dependence and the descriptions are included in the section \ref{WeatherDrivenParams}.
$T$ describes temperature.
}
\label{table:2}
\end{table}

\subsection{Weather driven mosquito parameters}\label{WeatherDrivenParams}
In this section we illustrate the weather-driven model parameters, forcing functions, transition functions between stages of the life cycle, and mortality functions to adapt the model to \textit{Culex} in the Cook county region. 
In our model, we neglect mosquito dispersal and our model is applicable only for an isolated habitat patch. 
Due to the significant phenotypic variability observed in the \textit{Culex} mosquito as well lack of information about the model parameters, we determine the parameter values using expert knowledge of local \textit{Culex} mosquito population biology and scientific literature \cite{CAILLY20127, EZANNO201539, YU201828, BHOWMICK2020110117, LAPERRIERE201199,  Jia, ijerph10051698}.
The two variables influencing the forcing functions are temperature ($T$) and precipitation ($P$), both subject to temporal fluctuations.
We use daily mean temperature and precipitation from $2014$ to $2019$ to perform the following simulations.
This study concentrates on aspects of \textit{Culex} mosquito biology where temperature exerts significant influence: aquatic development, mortality, maturity. 
The model is tailored to replicate the population dynamics of aquatic and adult \textit{Culex} over multiple season.
The progression between aquatic stages and the development of engorged adults is contingent upon temperature variations. 
The pace of development for \textit{Culex} stage $C$, impacted by temperature, can be mathematically formulated as a function of the temperature at a given time $t$, denoted as $T(t)$, thus incorporating the seasonality.
The Logan curves, customised for various species of \textit{Culex}, is applied to both the larval and pupal stages \cite{CAILLY20127, EZANNO201539, YU201828} to depict the correlation between temperature (in °C) and the rate of development across the range of $10$°C to $35$°C.
The mortality rates of larvae, pupae, and adults are derived from \cite{CAILLY20127, EZANNO201539, YU201828} and tailored to suit \textit{Culex} species in a temperate climate.
These rates are contingent upon temperature variations and we illustrate the functional relationship of all these different parameters.
For a thorough explanation of these parameters' derivation, please refer to the \cite{CAILLY20127}.
Eggs to Larvae maturity can be described by the following equation
\begin{equation}
\gamma_L = 0.16(e^{0.105(T-10)}-e^{0.105(35-10)-(35-T)/5.007})
\end{equation}
Similarly, after following the authors in \cite{CAILLY20127}, pupae to emerging adult maturity is defined as 
\begin{equation}
\gamma_A = 0.021(e^{0.162(T-10)}-e^{0.162(35-10)-(35-T)/5.007}),
\end{equation}
Larvae to pupae maturity ($\gamma_P$)  as $\gamma_A/4$, pupae mortality ($\beta_P$) as
\begin{equation}
\beta_P = e^{-T/2}+\mu_P,
\end{equation}
Larvae mortality ($\beta_L$)  as 
\begin{equation}
 \beta_L= 0.0025T^2-0.094T+1.0257,
\end{equation}
and the adult mosquito mortality $\beta_A$  as $\beta_L/10$  \cite{BHOWMICK2020110117, LAPERRIERE201199}.
We also adapt an established relationship between the egg ovipostion rate and temperature proposed by \cite{CAILLY20127, Jia, YU201828}
and define the egg laying rate ($\gamma_{A_{0}}$) as following:
\begin{equation}
\gamma_{A_{0}} = \max\left\{ { - 15.837 + 1.289T - 0.0163T^{2} , 0} \right\}
\end{equation}
The authors report that under non-extreme drought or rainfall conditions, \textit{Culex} abundance is more likely to be enhanced \cite{VALDEZ201728, Federico, Cole}.
Precipitation is recognised to influence the population growth of \textit{Culex} during the aquatic stages, thereby constraining its growth through flushing and flooding activities \cite{10.1603/ME10117}.
In our current study, we consider the role of precipitation impacting the environmental carrying capacity of aquatic stages (larvae and pupae), thereby increasing the number of available breeding sites.
Therefore, we adjust the environment’s carrying capacity ($K_C$) to incorporate the influence of rainfall as following:
$K_C (t) = \kappa_C \left(P_{\text{norm}}(t)+1\right),  C\in\{L,P\}$,
following the approach in \cite{CAILLY20127, EZANNO201539, ijerph10051698}, we define, $P_{\text{norm}}(t)$ as the rainfall amount accumulated over a two-week period, normalised to vary between $0$ and $1$.
Furthermore, we incorporate additional larval mortality ($\beta_W$) resulting from flushing whenever a total of $2.5$ cm of rain occurs over a $7$-day period.
We include the graphical illustrations of weather driven parameters in the Supplementary Information (SI).

\section{Basic offspring number $R_{0_{F}}$}\label{R0}
The fundamental parameter in population dynamics theory is the basic offspring number, $R_{0_{F}}$.
Depending on this value, the steady-state condition can either be a  trivial ($R_{0_{F}}\leq 1$) or, non-trivial ($R_{0_{F}}>1$) equilibrium point.
The model \eqref{Eq1} system has a trivial equilibrium or, mosquito free equilibrium point (MFE) $\mathcal{T}_0 = (E_{0}, L_{0}, P_{0}, A_{0}, A_{B_{0}}, A_{En_{0}}, A_{El_{0}}) = (0, 0, 0, 0, 0, 0, 0)$.
We provide now conditions for the extinction of the model system \eqref{Eq1}.
This basic offspring number, $R_{0_{F}}$ is the spectral radius of the next generation operator of the population \cite{YANG_MACORIS_GALVANI_ANDRIGHETTI_WANDERLEY_2009, doi:10.1142/S0218339015500278}.
With this purpose, we linearise the model system \eqref{Eq1} around the MFE.
After following the Next Generation Matrix method as described in \cite{doi:10.1098/rsif.2009.0386, VANDENDRIESSCHE200229}, we get the following matrices:

$\mathbb{F} =
\tiny{
\left[\begin{matrix}0 & 0 & 0 & 0 & 0 & 0 & \alpha_A \gamma_{A_{0}}\\0 & 0 & 0 & 0 & 0 & 0 & 0\\0 & 0 & 0 & 0 & 0 & 0 & 0\\0 & 0 & 0 & 0 & 0 & 0 & 0\\0 & 0 & 0 & 0 & 0 & 0 & 0\\0 & 0 & 0 & 0 & 0 & 0 & 0\\0 & 0 & 0 & 0 & 0 & 0 & 0\end{matrix}\right]}
$,\\
$\mathbb{V} =$\\
$ \tiny{
\left[\begin{matrix}- \beta_E - \gamma_L \psi & 0 & 0 & 0 & 0 & 0 & 0\\- \gamma_L \psi & \beta_L + \gamma_P + \frac{2 L \beta_{1}}{K_L} & 0 & 0 & 0 & 0 & 0\\0 & - \gamma_P & \beta_P + \gamma_A & 0 & 0 & 0 & 0\\0 & 0 & - \gamma_A \sigma_A e^{- \gamma_{em} \left(1 + \frac{P}{K_P}\right)} + \frac{P \gamma_A \gamma_{em} \sigma_A e^{- \gamma_{em} \left(1 + \frac{P}{K_P}\right)}}{K_P} & \beta_A + \gamma_B & 0 & 0 & 0\\0 & 0 & 0 & - \gamma_B & \beta_A + \gamma_{En} + \mu_B & 0 & 0\\0 & 0 & 0 & 0 & - \gamma_{En} & \beta_A + \gamma_{El} & 0\\0 & 0 & 0 & 0 & 0 & - \gamma_{El} & \beta_A + \gamma_{A_{0}}\end{matrix}\right]
}
$

\begin{equation}\label{R0BasicOffspring}
R_{0_{F}} = \rho(\mathbb{F}{\mathbb{V}}^{-1})
\end{equation}

\begin{eqnarray}\label{R011}
R_{0_{F}} = \left(\frac{\alpha_A \gamma_{A_{0}}}{\beta_E + \gamma_L \psi}\right)  
\left( \frac{\gamma_{L}}{\beta_L + \beta_W + \gamma_P} \right)  
\left( \frac{\gamma_P \sigma_A e^{- \gamma_{em}}}{\beta_P+\gamma_A} \right) 
\left( \frac{\gamma_A}{\beta_A+\gamma_B}\right) \\ \nonumber
\left( \frac{\gamma_B}{\beta_A+\gamma_{El}} \right)
\left(  \frac{\gamma_{En}} {\beta_A+\gamma_{En}+\mu_B}\right) 
\left(  \frac{\gamma_{El}} {\beta_A+\gamma_{A_{0}}}\right) 
\end{eqnarray}

Biological meaning of the terms in \eqref{R011} can be interpreted immediately.
The terms in the expression of $R_{0_{F}}$ during the non-diapausing phase, can ecologically be illustrated as the product of fraction of eggs laid by the egg laying mosquitoes mature into larvae stage in aquatic stage, 
the number of potential larvae that become pupae after facing the water flush, the number of surviving pupae maturing into emerging adult mosquitoes in terrestrial stage, the number of 
emerging adult mosquitoes maturing into blood seeking mosquitoes, the number of potential blood seeking mosquitoes that become engorged mosquitoes and the number of engorged mosquitoes maturing into egg laying mosquitoes.
Thus, the threshold quantity $R_{0_{F}}$ in \eqref{R011} can be quantified as the average expected number of new adult female offsprings produced by a single female mosquito during its lifespan.

\section{Model simulation}\label{Modelsimulation}
The model predicts the abundance of mosquitoes per stage $(E, L, P, A, A_{B}, A_{En}, A_{El})$ over time. 
In addition, we aggregate the dynamic output computed by the model \eqref{Eq1} after following the authors in \cite{CAILLY20127, EZANNO201539, YU201828} and we calculate the average the daily number of blood-seeking adults ($A_{B}$), emerging adults ($A$) and Egg laying adults ($A_{El}$).
We aggregate these outputs for their epidemiological importance as we utilise them to compare the  relative abundance entomological collections and the predictions of the model as describe in the section \ref{Modelvalidation}.\par
We download weather data from the PRISM Climate Group \cite{PRISMClimateGroup} for the period spanning 2004 to 2019.
After following the authors in \cite{CAILLY20127, EZANNO201539, YU201828, BHOWMICK2020110117}, in our R \cite{R} implementation, we execute the model \eqref{Eq1} and \eqref{Eq1Modified}, conduct simulations spanning $10$ consecutive years to achieve a stationary state. 
This ensures that external weather-driven forcing functions stabilise, reaching an equilibrium state and thus minimising undesired fluctuations in the model outputs. 
At the onset of our simulations, corresponding to the $1{\text{st}}$ of January, the mosquito population comprises solely the overwintering stage.\par
We employ an implicit Euler scheme with a daily time step to numerically solve the model system described by \eqref{Eq1} and \eqref{Eq1Modified}. 
This time step aligns with the intervals for which the weather data are accessible, as outlined in the section \ref{Weatherdata}.
Initially, the population comprises $10^6$ eggs and $10^2$ emerging, blood-seeking adult mosquitoes, starting from the 1st of January \cite{CAILLY20127} . 
Following the methodology outlined in \cite{BHOWMICK2020110117, LAPERRIERE201199}, we maintain the number of overwintering mosquitoes ($N_{\text{Min}}$) at $1\%$ of the initial condition in subsequent seasons in our simulations. 
We adopt this approach to prevent the extinction of the mosquito population during simulations, as demonstrated by the authors in \cite{BHOWMICK2020110117, LAPERRIERE201199}. 
Further details regarding the initial conditions are provided in the Supplementary Information.

\section{Evaluating the impact of temperature on mosquito population size}\label{TvsMosquito}
We analyse the ecological and epidemiological findings derived from the mathematical modelling previously discussed are now being correlated with the entomological parameters determined from temperature-controlled experiments outlined in the preceding section.
\begin{figure}[H]
\centering
\includegraphics[width=10cm]{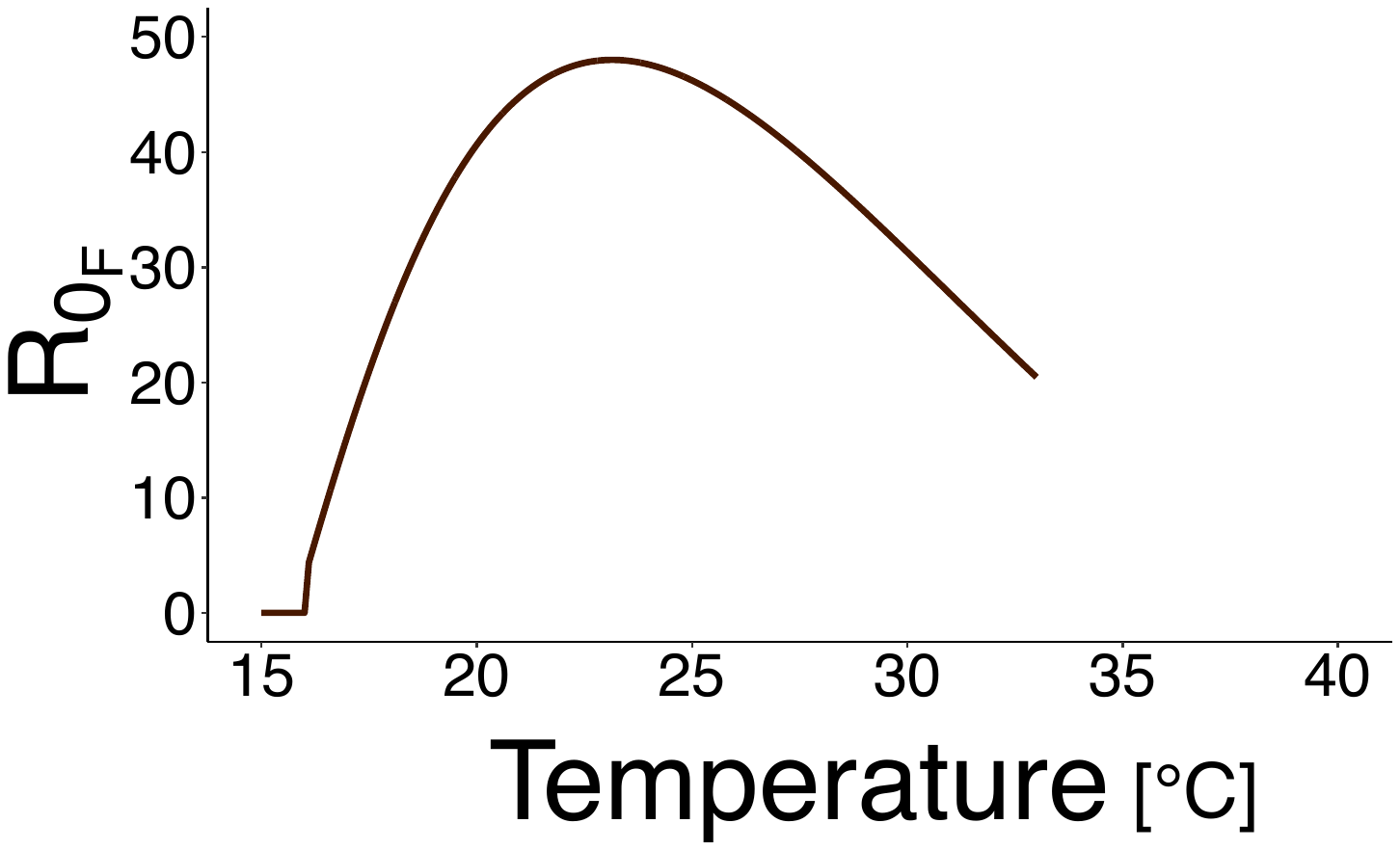}
	\caption{ The basic offspring number $R_{0_{F}}$ as a function of temperature is shown. 
	The model presupposes the life-cycle of \textit{Culex} as depicted in the flowchart illustrated in Figure \ref{fig:ModelFlow}.
	}
    \label{fig:R0FvsTemp}
\end{figure}

Figure \ref{fig:R0FvsTemp} shows the curve of the Basic Offspring number $R_{0_{F}}$, using \eqref{R011}.
We can observe from the section \ref{R0} that the mosquito population is unsustainable in a habitat patch if  $R_{0_{F}}<1$.
The equation \eqref{R011} incorporates temperature-dependent entomological parameters, and the input for these parameters is provided by the downloaded temperature data from PRISM for the NWMAD region.
From the figure \ref{fig:R0FvsTemp}, it is clear that $R_{0_{F}}$ is greater than the unity when the temperature is more than $16^{\circ} C$.
Within this temperature range, there may be no transmission of WNV as the region is potentially devoid of mosquitoes and when $R_{0_{F}}>1$, which occurs in the temperature increases from  $16^{\circ} C$. 
When $R_{0_{F}}>1$, occurring within the temperature range of $20^{\circ}C<T<26^{\circ}C$, there is a heightened risk of WNV outbreaks.
It's important to emphasise that this range represents a higher likelihood of WNV transmission based solely on the magnitude of $R_{0_{F}}$.
However, WNV incidence is not solely determined by $R_{0_{F}}$ but also by factors such as the availability of birds to amplify the transmission cycle of WNV, the susceptibility of mosquitoes to acquire the infection, and the frequency of contacts between them.

\section{Assessing different abatement strategies}\label{ImpactofZetaOnMosquito}
Mosquito abatement includes various strategies to control mosquito populations \cite{EZANNO201539, Demers, 10.1093/jme/tjad088, 10.2987/19-6848.1}. 
Spraying adulticides is a common strategy to control adult mosquito populations and reduce the risk of mosquito-borne diseases. 
The application of ultra-low volume (ULV) aerial sprays containing organophosphate or pyrethroid insecticides has demonstrated effectiveness in controlling adult mosquito populations, potentially thwarting WNV outbreaks by diminishing the vector population \cite{Demers, 10.1093/jme/tjad088, 10.2987/19-6848.1, 10.2987/19-6848.1}.
Formulating the functional form of $\zeta$ (rate of effectiveness of adulticide that yields mosquito mortality rate) holds significance, mentioned in the section \ref{Mathematicalmodel}.
Given that ULV application occurs exclusively during the summer \cite{10.1093/jme/tjad088, EZANNO201539}, we opt for a straightforward step function approach, as suggested in \cite{BHOWMICK2024107346}. 
The step function $\zeta(t)$ delineates the mortality rate of mosquitoes consequent to adulticide treatment, signifying a proportional reduction.
We assess the effectiveness of various adulticide spraying approaches within a single season and the impact of different scales of efficacy on $R_{0_{F}}$.
\begin{equation}
\zeta (t) =
 \begin{cases} 
     \zeta_0 & t^{\mbox{apply}}\leq t\leq t^{\mbox{apply}}+ t^{\mbox{duration}} \\
      0 & \mbox{Otherwise} 
   \end{cases}
\end{equation}
Here $ t^{\mbox{apply}}$ is the day of the ULV treatment application,  $t^{\mbox{duration}}$ is the duration of the treatment in days and 
$\zeta_0$ is the the daily ULV treatment effectiveness.

\subsection{Assessing the effect of ULV spray on relative mosquito abundance}\label{ULVSpray1}
\begin{figure}[H]
\centering
\includegraphics[width=12cm]{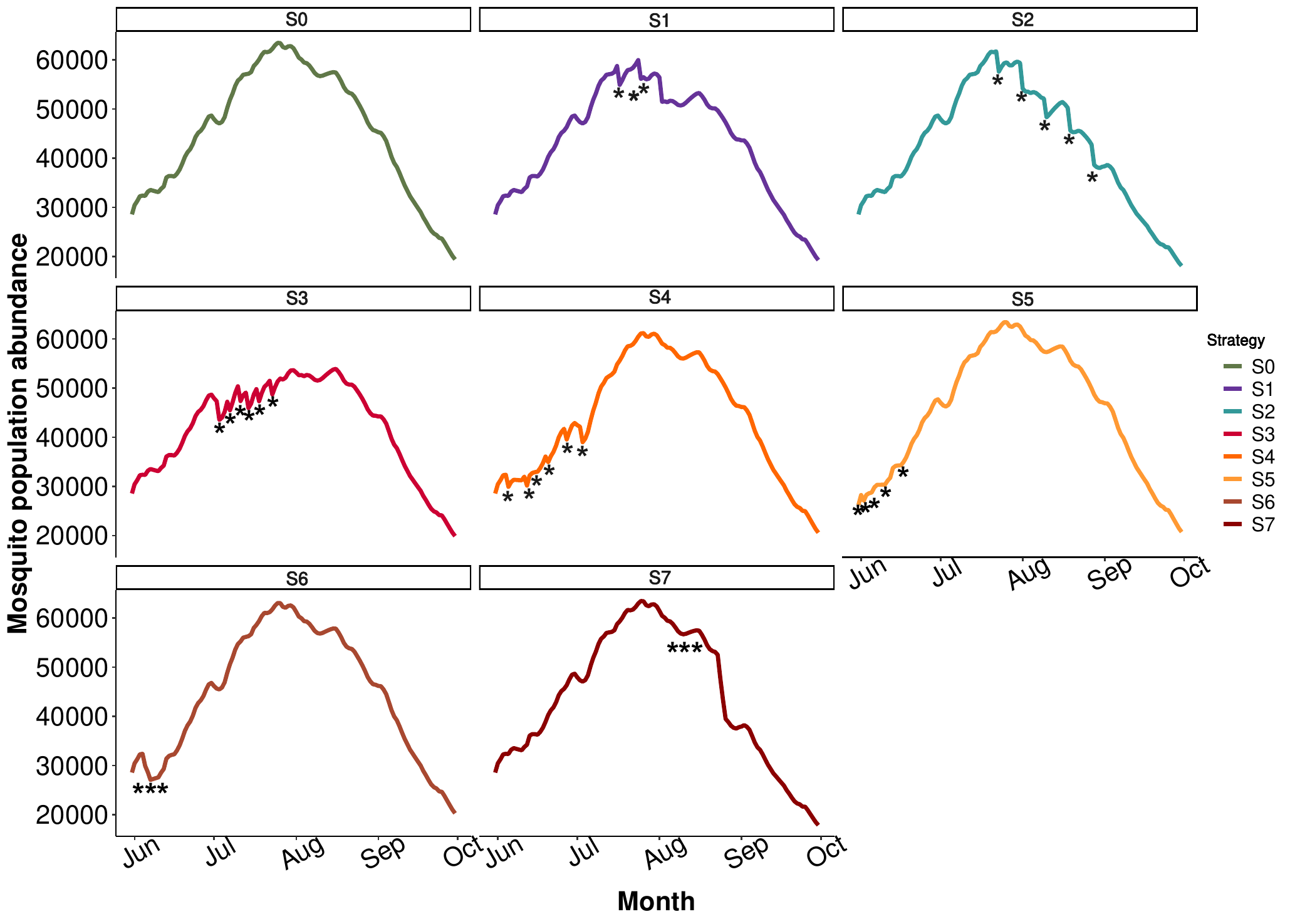}
	\caption{ 
 We display the relative abundance of mosquitoes observed throughout the simulation period under different control strategies.
 The stars indicate the days when adulticide is sprayed, as per the schedule outlined in the table \ref{table:A}.
           }
    \label{fig:abatement2}
\end{figure}

\begin{figure}[H]
\centering
\includegraphics[width=12cm]{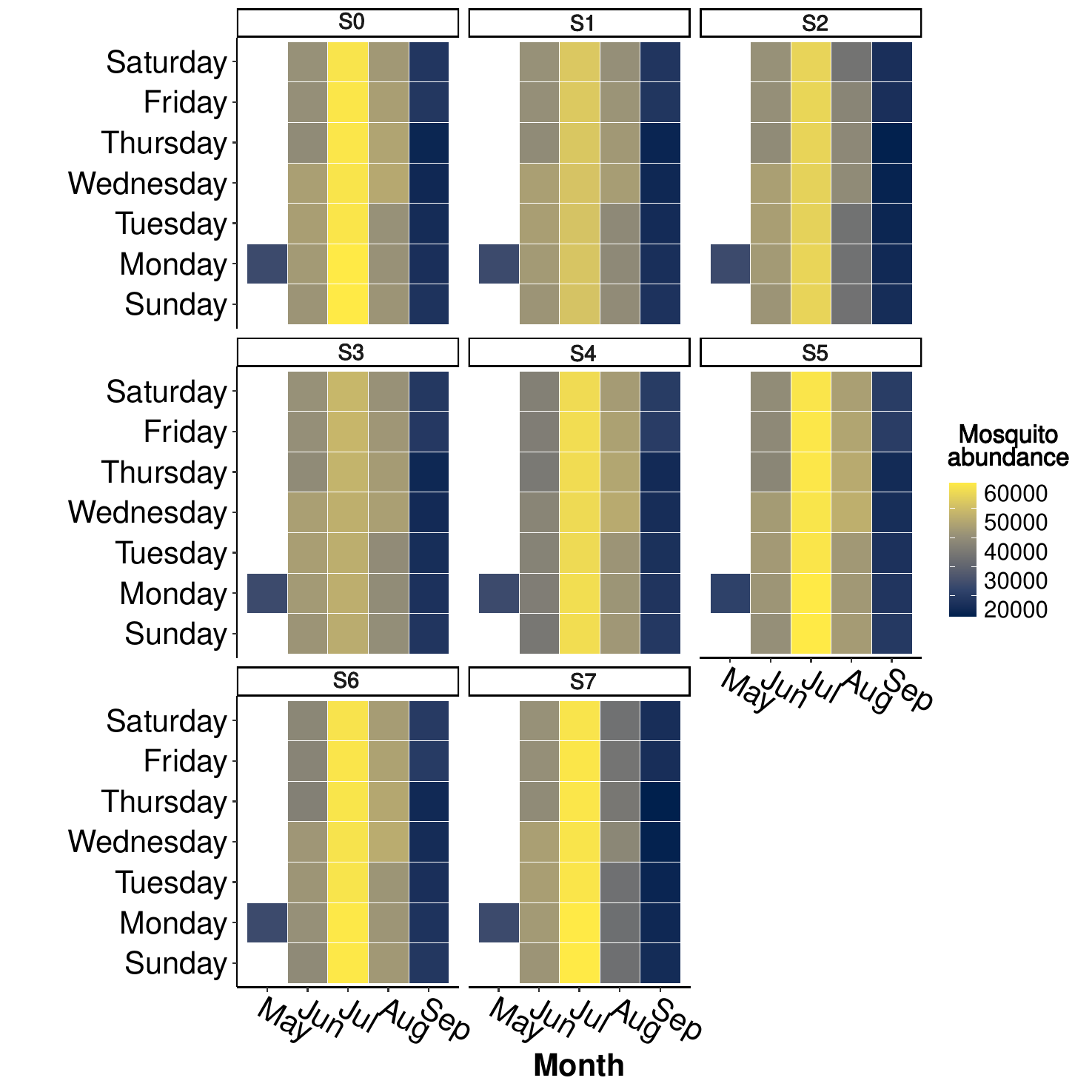}
	\caption{ 
Daily mosquito abundance is cumulatively simulated and presented in a calendar plot, which includes the days when adulticide is applied, as listed in the table \ref{table:A} and shown in the  figure \ref{fig:abatement2}.
           }
    \label{fig:abatement1}
\end{figure}

Several strategies and techniques can be employed to apply adulticides effectively \cite{10.2987/19-6848.1, 10.1093/jme/tjz083, EZANNO201539, Demers}.
Mathematical modelling and simulations can play an important role in planning and implementing effective mosquito abatement strategies, including the application of adulticides.
NWMAD employs various strategies, and we have simulated some strategies based on expert knowledge to compare them \cite{10.2987/19-6848.1, 10.1093/jme/tjz083}.
We evaluate seven distinct spraying strategies and compare them to a scenario where no adulticide is being sprayed through simulations.
The references for these strategies are denoted as $S_i$ (where $i$ ranges from $1$ to $7$), with $S_0$ representing the scenario where no adulticide is being sprayed.
In our simulations, we simulate under the assumption that the adulticide primarily impacts emerging adult mosquitoes, those actively seeking blood meals and looking for refuge to lay eggs.
We assume the value of daily ULV treatment effectiveness is $60\%$ while performing our simulations and this results in an additional mortality factor affecting both the emerging and blood-seeking mosquito populations.
As a consequence, the model equations \eqref{Eq1} as mentioned in the section \ref{Mathematicalmodel} governing these two compartments undergo modification to reflect this additional mortality factor and the modified equations change to:
\begin{eqnarray}\label{Eq1Modified}
\frac{dA}{dt} &=& \gamma_A\sigma_A Pe^{-\gamma_{em}(1+\frac{P}{K_P})}-\beta_A A-\gamma_{B}A-\zeta A\\ \nonumber
\frac{dA_B}{dt}&=& \gamma_{B}A-\beta_A A_B-\mu_B A_B-\gamma_{En}A_B-\zeta A_B\\ \nonumber
\frac{dA_{El}}{dt} &=&\gamma_{El}A_{En}-\beta_A A_{El}-\zeta A_{El}\\ \nonumber
\end{eqnarray}
Figures \ref{fig:abatement1} and  \ref{fig:abatement2} represent the simulated abundance of \textit{Culex} population under the different scenarios and we use the weather data from $2014$. 

\begin{table}[H]
\centering
\begin{tabular}{||c c c c ||} 
 \hline
 Spray start time & Interval  & Number of times & Strategy name \\ [0.5ex] 
 \hline\hline
No spray & 0 & 0 & $S_0$   \\ 
Mid July &  Once a week & $3$ weeks & $S_1$\\
Late July &  Once a week & $5$ weeks & $S_2$\\
First week of July &  Twice a week & $3$ weeks & $S_3$\\ 
First week of June &  Once a week & $5$ weeks & $S_4$\\
Mid May&  Once a week & $5$ weeks & $S_5$\\
First week of June &  Everyday & $3$ consecutive days & $S_6$\\
Third week of August &  Everyday & $3$ consecutive days & $S_7$\\
[1ex] 
 \hline
\end{tabular}
\caption{
NWMAD prescribed various spray regimes employed in our model \eqref{Eq1},  \eqref{Eq1Modified} to assess optimal strategies.
}
\label{table:A}
\end{table}


\subsection{Assessing the effect of ULV spray on \textit{Basic Offspring Number} }
\begin{figure}[H]
\centering
\includegraphics[width=12cm]{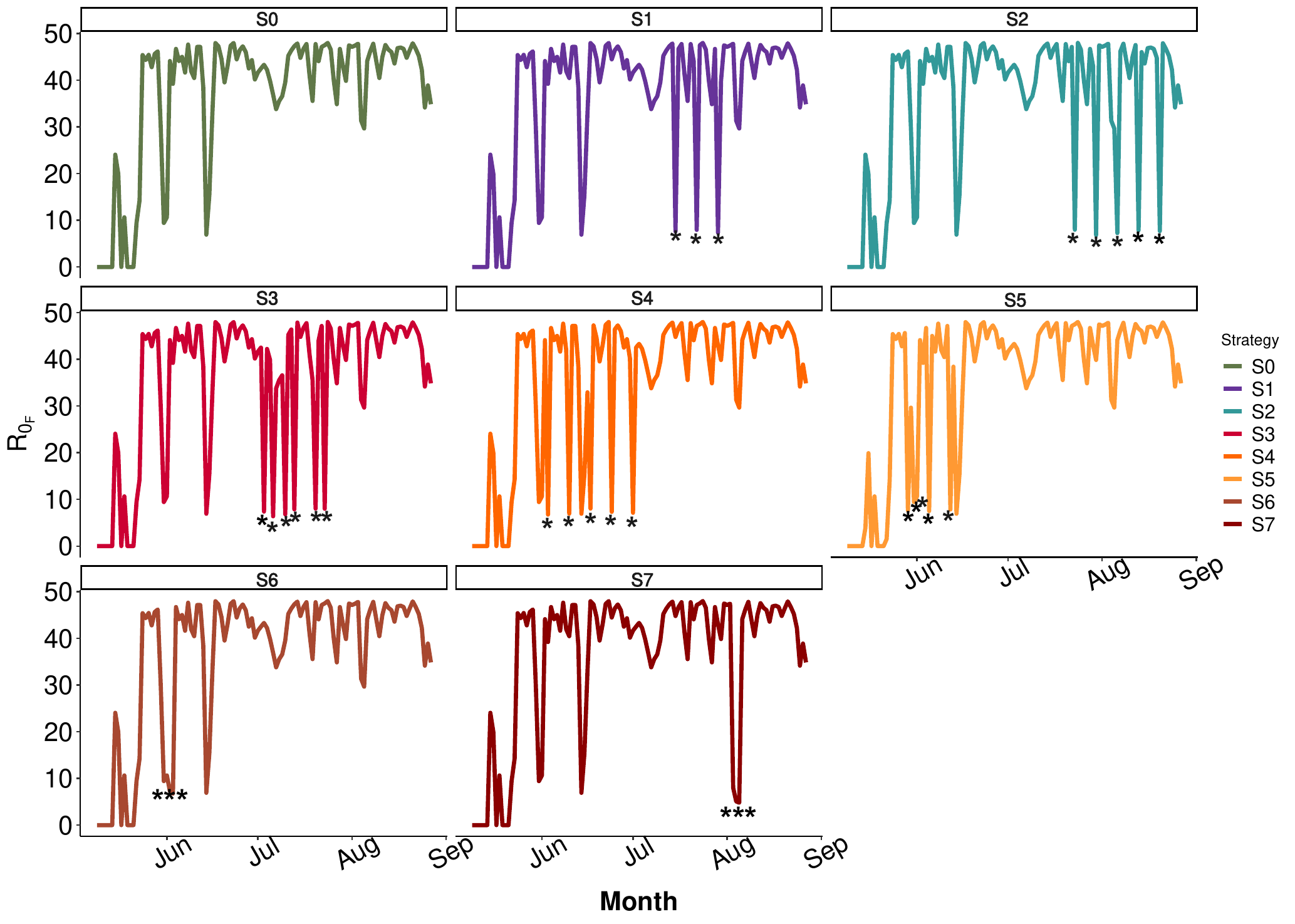}
	\caption{ 
 We present the basic offspring number $R_{0_{F}}$ observed over the simulation period for different control strategies. 
 Stars mark the days when adulticide is  applied, following the schedule outlined in table \ref{table:A}.
           }
    \label{fig:abatement2R0}
\end{figure}

\begin{figure}[H]
\centering
\includegraphics[width=12cm]{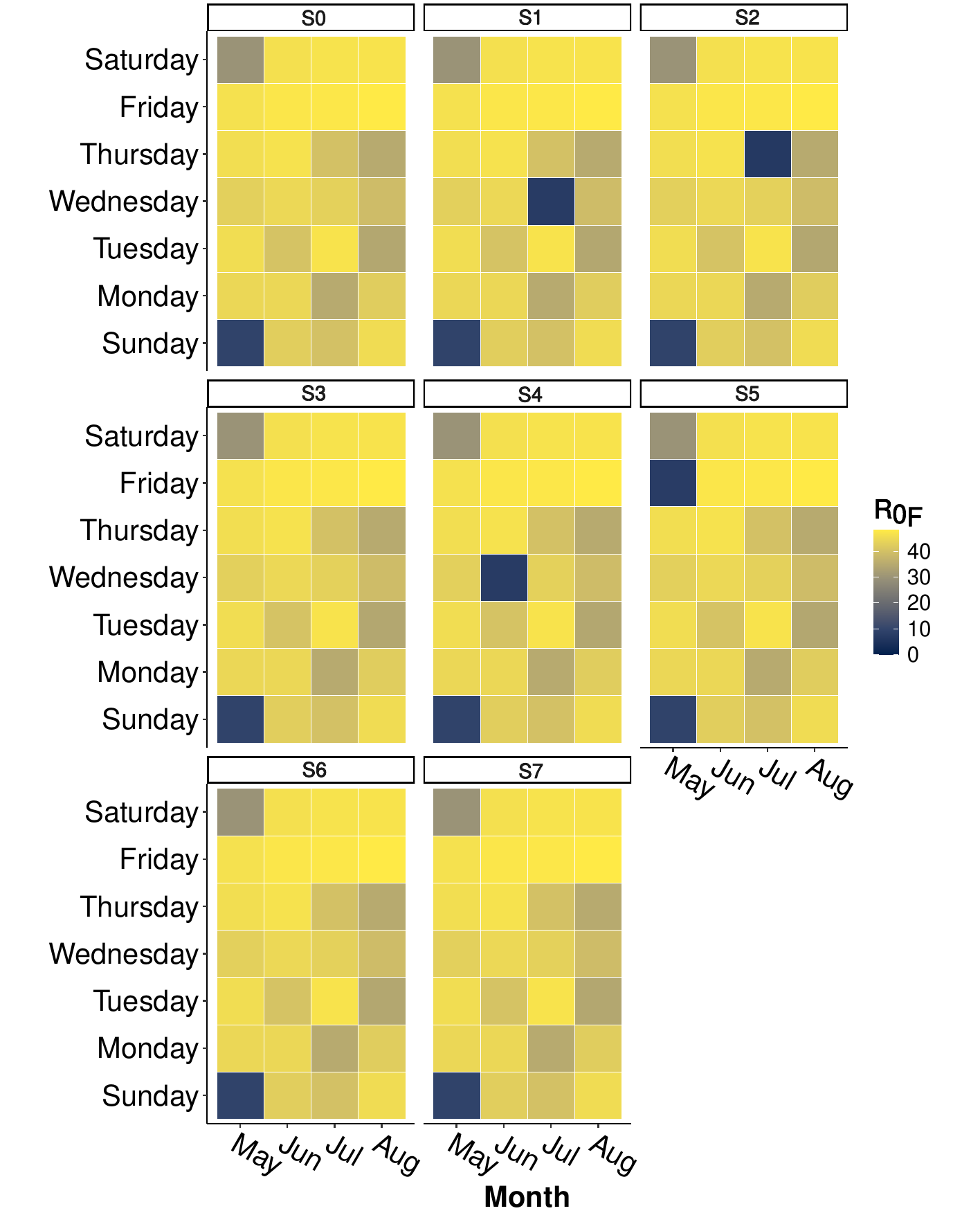}
	\caption{ 
This figure illustrates the basic offspring number, $R_{0_{F}}$, under various control strategies. 
The values of $R_{0_{F}}$ are cumulatively simulated per day and displayed in a calendar plot, which also marks the days when adulticide is applied, as indicated in the table \ref{table:A} and depicted in the figure  \ref{fig:abatement2R0}.
           }
    \label{fig:abatement1R0}
\end{figure}

If the basic reproduction number $R_{0_{F}}$ can be reduced to a value below one through ULV spraying, then theoretically, the incidence of WNV cases can be decreased.
It's worth noting that achieving this outcome is not feasible.
When we include the impact of adulticide, then  the expression of basic offspring number ($R_{0_{F}}$) changes into the following:
\begin{eqnarray}\label{R01Ad}
R_{0_{F}} = \left(\frac{\alpha_A \gamma_{A_{0}}}{\beta_E + \gamma_L \psi}\right)  
\left( \frac{\gamma_{L}}{\beta_L + \beta_W + \gamma_P} \right)  
\left( \frac{\gamma_P \sigma_A e^{- \gamma_{em}}}{\beta_P+\gamma_A} \right) 
\left( \frac{\gamma_A}{\beta_A+\gamma_B+ \zeta}\right) \\ \nonumber
\left( \frac{\gamma_B}{\beta_A+\gamma_{El}} \right)
\left(  \frac{\gamma_{En}} {\beta_A+\gamma_{En}+\mu_B+ \zeta}\right) 
\left(  \frac{\gamma_{El}} {\beta_A+\gamma_{A_{0}}+ \zeta}\right) 
\end{eqnarray}
The biological significance of the terms in equation \eqref{R01Ad} can also be elucidated similarly to what we did in equation \eqref{R011}, but this time incorporating the impact of adulticide as we describe in the section \ref{R0}.
Figures \ref{fig:abatement1R0} and  \ref{fig:abatement2R0} represent the simulated magnitude of $R_{0_{F}}$ under the different ULV spraying strategies as illustrated in the section
\ref{ULVSpray1} and we utilise the weather data from $2014$ for our simulations. 

\section {Sensitivity analysis}\label{SensitivityAnalysis}
\begin{figure}[H]
\centering
\subfloat[Sobol indeces]{\includegraphics[width=0.5\textwidth]{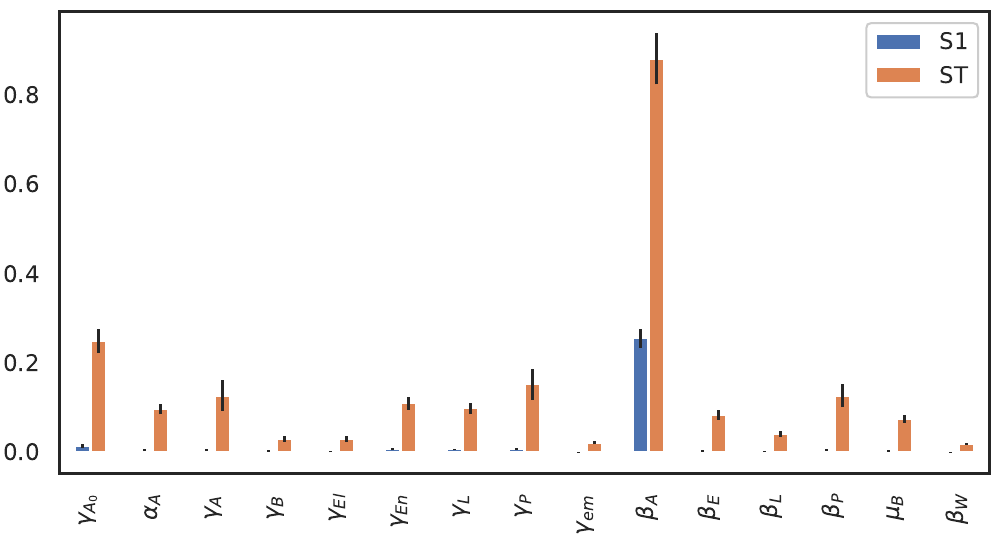}\label{fig:sen1}}
\hfill
\subfloat[Second-order interactions]{\includegraphics[width=0.45\textwidth]{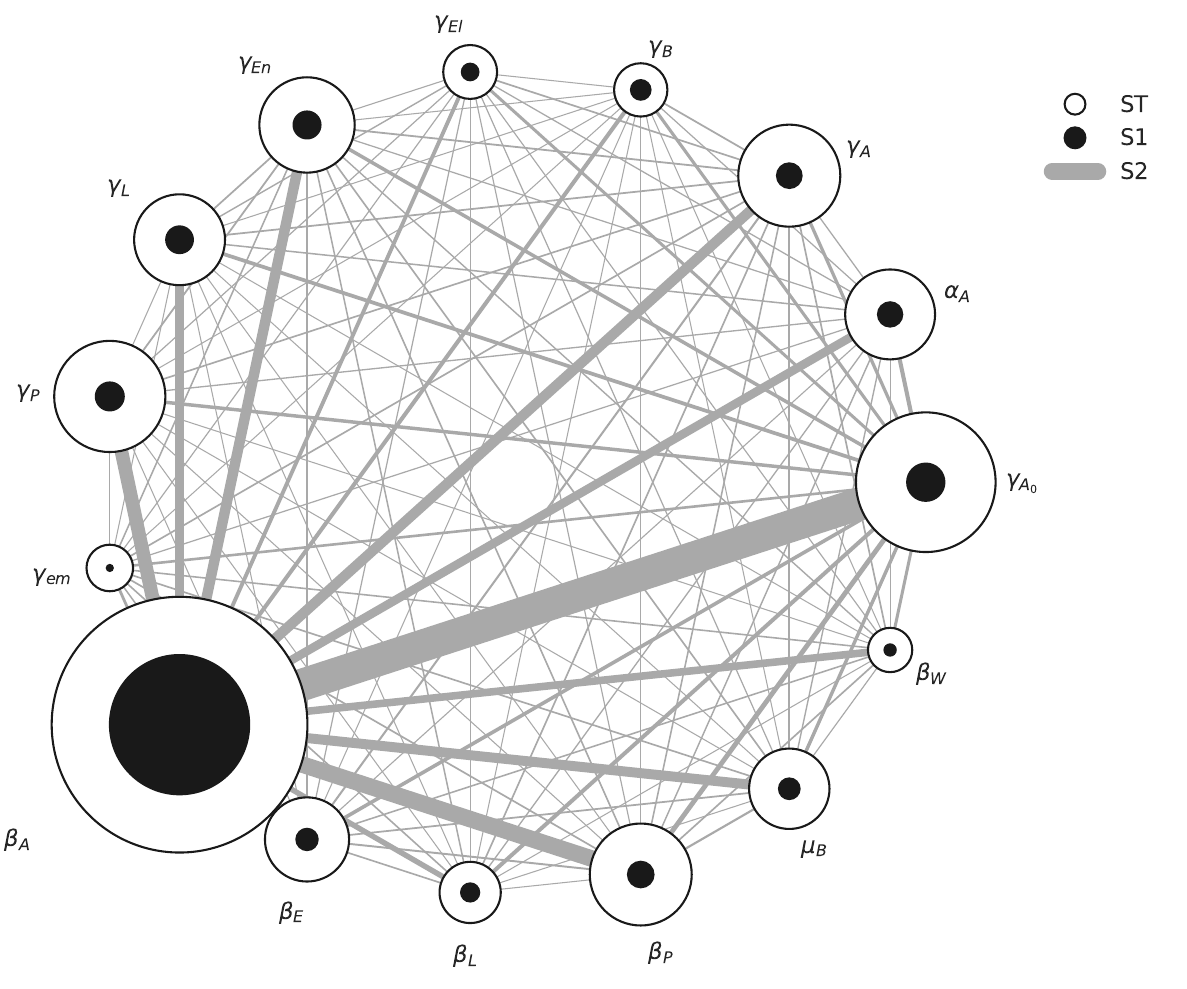}\label{fig:sens2}}
\caption{Sobol indices}\label{fig:Sensitvity}
\end{figure}

By utilising sensitivity analysis, we gain insights into the significance of different parameters, particularly in light of the variability in the basic offspring number $R_{0_{F}}$. 
Our emphasis lies on the global sensitivity analysis (GSA), which investigates into how model output variables respond to changes in model parameters.
We employ SALib \cite{Iwanaga2022, Herman2017} to calculate first-order ($S1$), second-order ($S2$), and total ($ST$) Sobol indices as depicted in the figure \ref{fig:Sensitvity}. 
These indices help to estimate the contribution of each input to output variance and identify input interactions, all with respect to the Basic Offspring number $R_{0_{F}}$ (using the mean value of $R_{0_{F}}$).
By default, $95\%$ confidence intervals are provided for each index.
We first select and visualise the total and first-order indices for each input and figure \ref{fig:sen1} reveals that $\beta_A$  has the highest $ST$ index and it indicates that it contributes over $90\%$ of $R_{0_{F}}$, variance while accounting for interactions with other parameters, followed by $\gamma_{A_{0}}$, $\gamma_A$, $\gamma_P$, $\beta_P$, $\alpha_A$.

It's necessary to mention that, especially concerning the $S1$ index, it signifies the individual contribution of each input to variance.
According to the Sobol sensitivity index, $Si$, where $i=1,2$, 
$\beta_A$, $\gamma_{A_{0}}$, $\alpha_A$ are identified as the primary contributors to the variance of the Basic offspring number ($R_{0_{F}}$).
We utilise a more advanced visualisation method to integrate the second-order interactions amongst the inputs, which are estimated from the $S2$ values.
Here, the sizes of the $ST$ and $S1$ circles represent the normalised importance of the variables associated with the response function.
The figure \ref{fig:sens2}, $\beta_A$ demonstrates strong interactions with $\gamma_{A_{0}}$ and $\beta_i$, where $i = E, L, P$ followed by the interactions amongst $\gamma_i$, where $i = L, P, A, En$, and $\alpha_A$.
From the figure \ref{fig:sen1}, we find that $R_{0_{F}}$ is mostly sensitive to the natural mortality of adult mosquito, followed by the egg laying rate, developmental rate of larvae to pupae, developmental rate of egg to larvae, developmental rate of pupae to adult, mortality rate of pupae and number of eggs laid by egg laying mosquito respectively.
Figure \ref{fig:sens2} indicates that natural mortality rate of adult mosquito has strong interactions with the egg laying rate, mortality rate of pupae, developmental rate of larvae to pupae, developmental rate of egg to larvae, developmental rate of pupae to adult, developmental rate of blood seeking mosquito into engorged and number of eggs laid by egg laying mosquito, respectively.
Figure \ref{fig:Sensitvity} illustrates several key messages that should be of concern within the context of mosquitoes lifecycle.
It gives us information about the contributions of the parameters to the variance of the basic offspring number $R_{0_{F}}$, with the total effect index, $ST$, encompassing both individual contributions and interaction effects.
The parameter distributions used to calculate the Sobol indices index is presented in the Supplementary Information (SI).

\section {Model Validation}\label{Modelvalidation}
In our current study, we embark on the crucial task of validating our weather-driven mechanistic ODE based mathematical model against real-world mosquito trap data acquired from NWMAD.
Accurate model validation is essential for ensuring the reliability and predictive capability of our mathematical model in replicating the observed abundance of \textit{Culex} population. 
By comparing the simulated model outputs with empirical data collected from mosquito traps, we aim to assess the model's ability to  represent the intricacies of mosquito abundance over time. 
This validation process can serve as a critical step in bolstering the credibility of our mathematical model and its potential applications in guiding effective mosquito control strategies under different scenarios.

\begin{figure}[H]
\centering
\subfloat[Model vs trap1 data]{\includegraphics[width=0.45\textwidth]{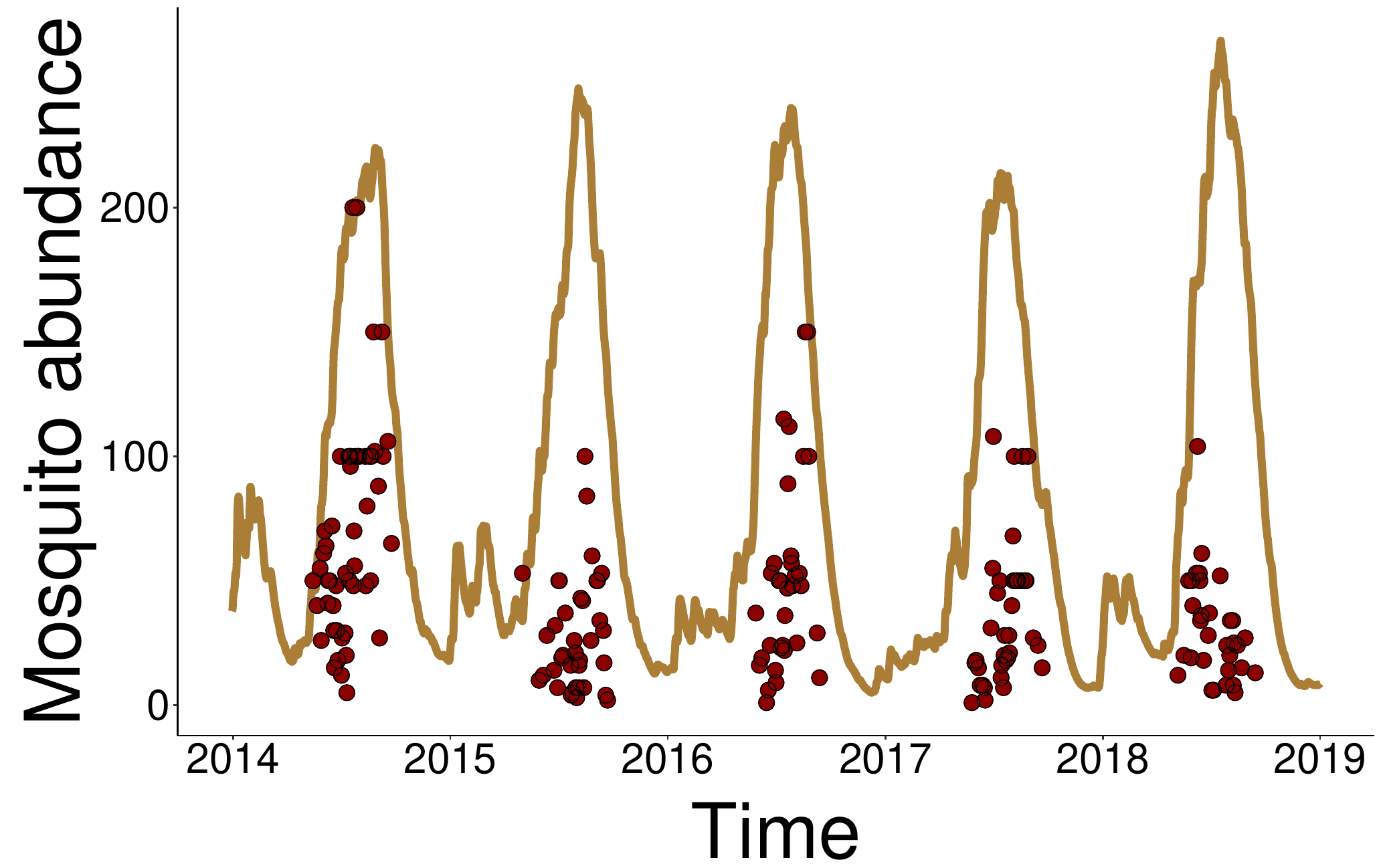}\label{fig:f1}}
\hfill
\subfloat[Model vs trap2 data]{\includegraphics[width=0.45\textwidth]{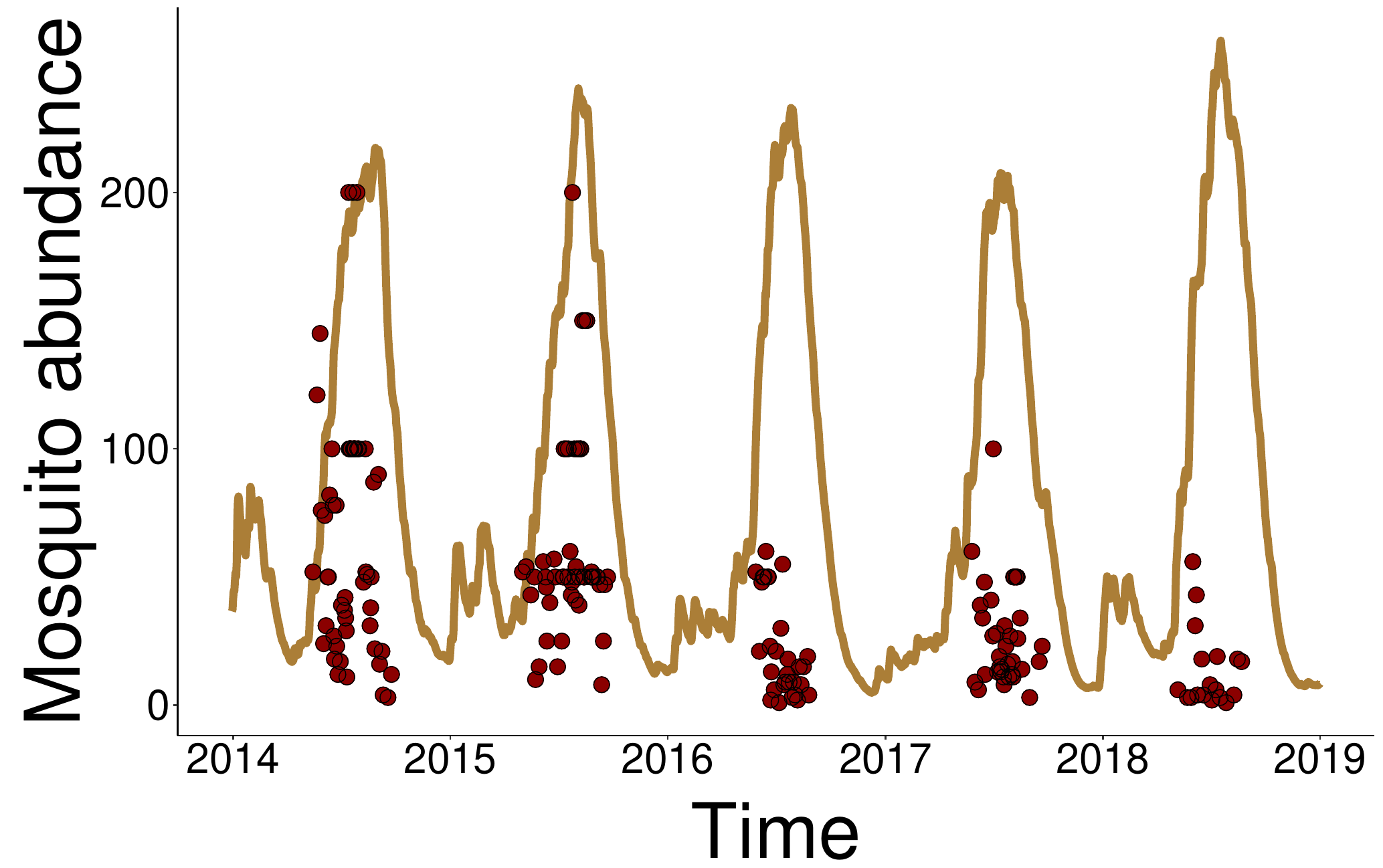}\label{fig:f2}}\\
\subfloat[Model vs trap3 data]{\includegraphics[width=0.45\textwidth]{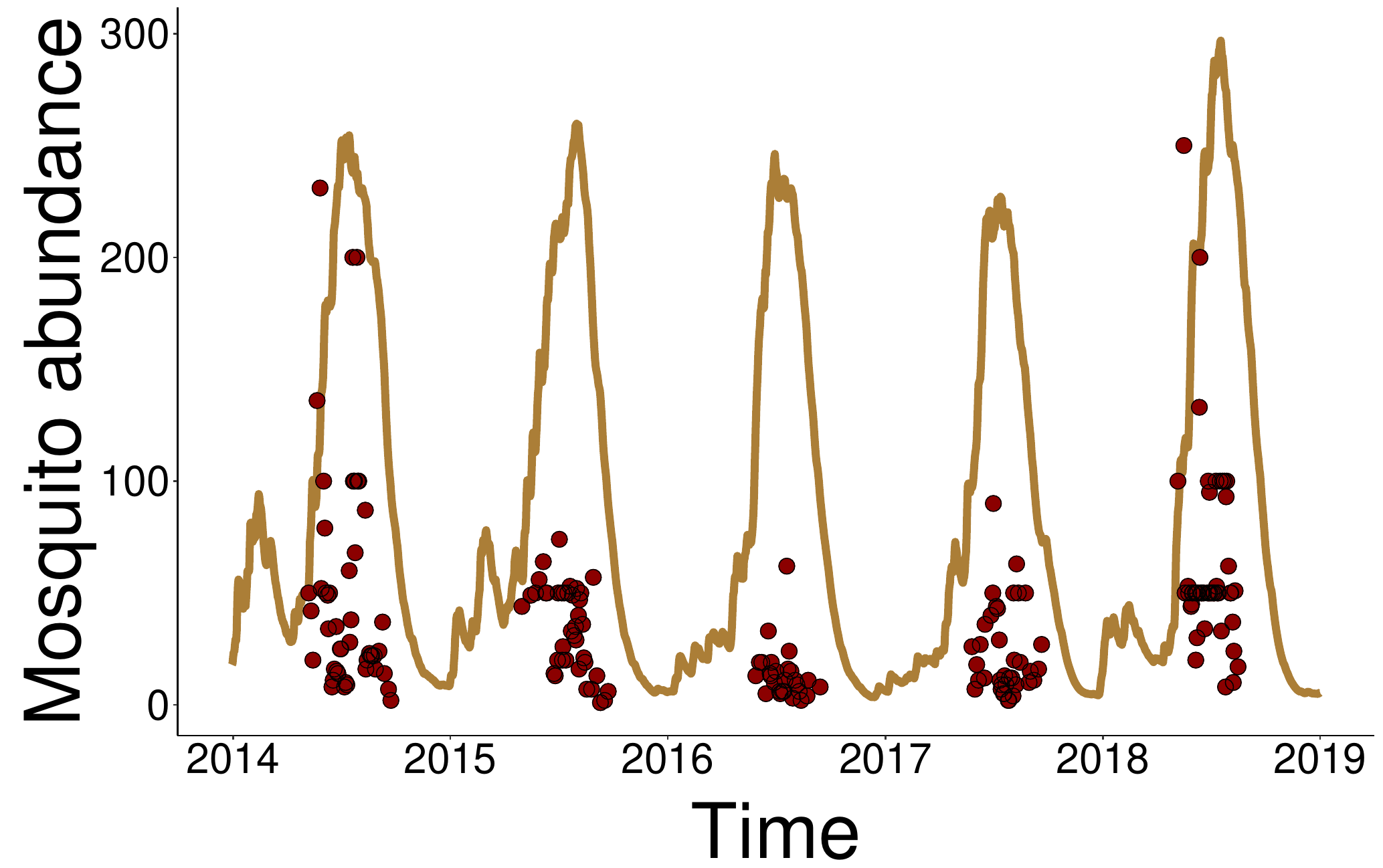}\label{fig:f3}}
\hfill
\subfloat[Model vs trap4 data]{\includegraphics[width=0.45\textwidth]{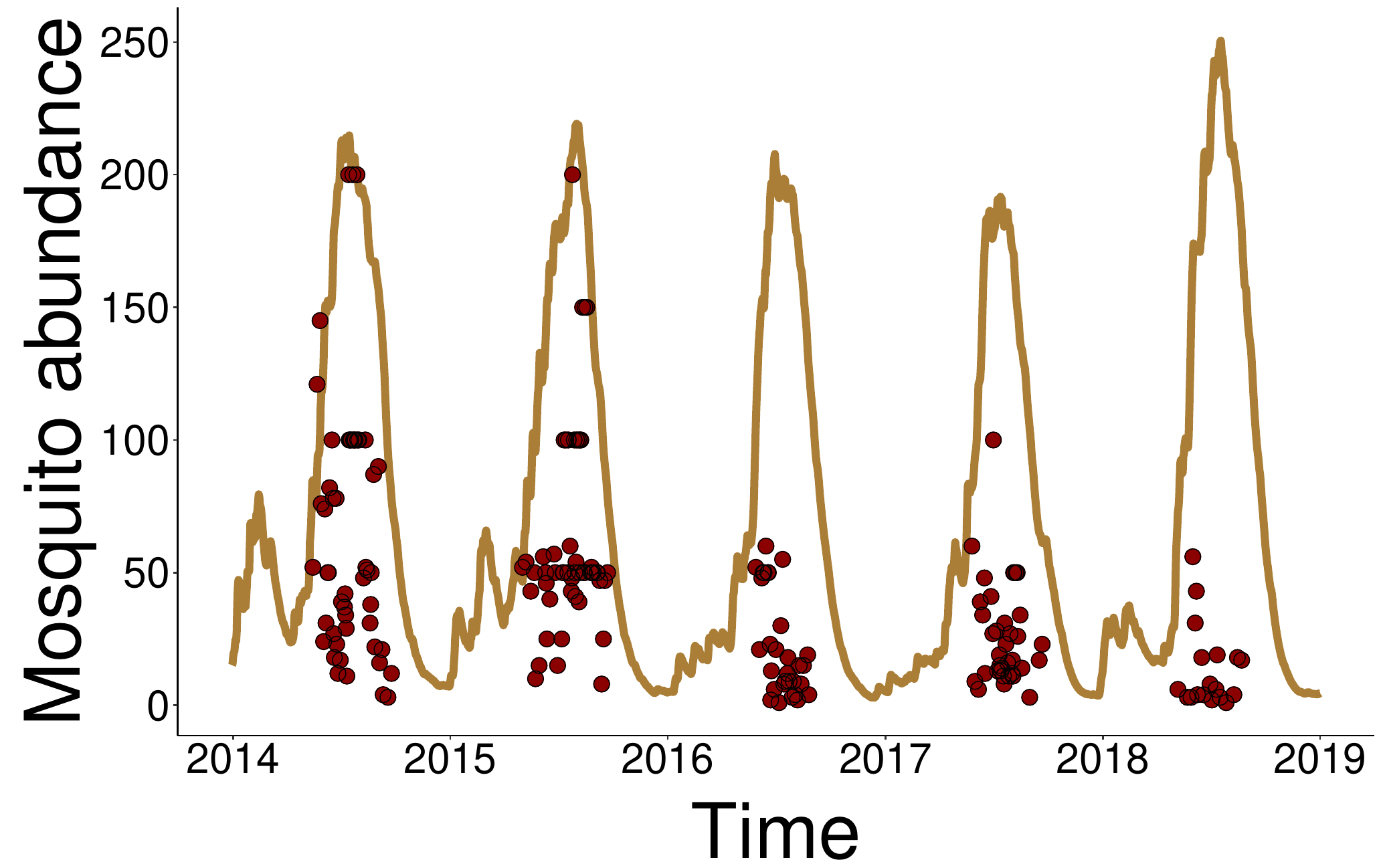}\label{fig:f4}}
\caption{A comparison between the model simulation generated from our model as described in \eqref{Eq1} and trap data. 
Here, we show the calibrated simulations with four different trap data sets for the demonstration purpose.
The red dots are the collected trap data and the yellow lines are the calibrated solution from the model system.
}\label{fig:ModelValidation}
\end{figure}

Figures \ref{fig:f1} (cross-correlation of $0.89$), \ref{fig:f2} (cross-correlation of $0.83$), \ref{fig:f3} (cross-correlation of $0.78$), \ref{fig:f4} (cross-correlation of $0.81$) display the seasonal fittings of model \eqref{Eq1} and \eqref{Eq1Modified} alongside the data from mosquito traps.
Different scaling factors and parameters are used for each trap location due to the deterministic nature of the model and yearly fluctuations in the data.
Our weather-driven mathematical model effectively reproduces the peak magnitude for most mosquito seasons of relative mosquito abundance.
However, it may encounter difficulty in capturing peaks with higher magnitudes or additional nonlinear characteristics, as this would necessitate a sharper derivative of the differential equation, potentially leading to stability issues with the numerical solver.
Furthermore, the model may struggle to depict the entire \textit{Culex} population dynamics and all peaks with lower magnitudes from the trap data, or additional nonlinear characteristics, as it requires weather-driven parameters and may face stability issues due to the discretisation of our ODE based mathematical model.
Additionally, the lack of information about the number of hibernating mosquitoes could contribute to the discrepancies.
The discrepancy between model predictions and observed data can be attributed to the simplified assumptions used in constructing our weather-driven mathematical model, thus warranting further investigation.
The goal of this model validation process is to evaluate how closely the predicted relative abundance of Culex mosquitoes from the weather-driven model matches the pooled data collected from various traps.
Our study uses NWMAD trap data and PRISM weather data from Cook County, Illinois, USA, covering the period from $2014$ to $2019$.
We validate our weather-driven mechanistic model \eqref{Eq1} and \eqref{Eq1Modified} with these datasets, highlighting its strength and robustness.
Unfortunately, data from other mosquito abatement districts are not readily available for validation or comparison. 
However, due to the flexibility and comprehensive design of our mathematical model, it can be extended to other regions when such data becomes accessible.
Nevertheless, the weather-driven model simulation generally provides time-series values of the \textit{Culex} population.

\section{Discussion and Conclusion}\label{Discussion}

In our research, we integrate two weather-associated variables  temperature and rainfall to parameterise our mechanistic model \ref{Eq1}. 
This model encompasses the entire mosquito life cycle and also incorporates the diapause mechanism in a straightforward manner. 
To the best of our knowledge, our model represents the first mechanistic approach to understanding the population dynamics of \textit{Culex} mosquitoes in a temperate climate, providing a closed-form expression for the Basic Offspring number ($R_{0_{F}}$) and its relationship with temperature. 
Our simulations align closely with data on the number of mosquito pools trapped, affirming the consistency and accuracy of our model as described in the section \ref{Modelvalidation}.

Ultimately, our simulations based study aim to inform evidence-based decision-making for public health authorities and policymakers engaged in the ongoing battle against vector-borne illnesses.
Furthermore, the research seeks to bridge the gap between theoretical modelling and practical implementation by incorporating the model into a comprehensive mosquito abatement planning framework as explained in the section \ref{ImpactofZetaOnMosquito}. 
The integration of real-time weather data into the planning process enables a dynamic and adaptive approach, thus optimising the allocation of resources and interventions based on current and forecasted meteorological conditions. 
By doing so, this study aims to provide a valuable tool for public health authorities and mosquito control agencies to enhance their ability to mitigate the spread of mosquito-borne diseases effectively.

\par
One important aspect of our model is its ability to simulate the impact of various vector control interventions (see the section \ref{ImpactofZetaOnMosquito}). 
By incorporating parameters related to adulticide spraying practices, our weather-driven model can perform the evaluation of different intervention strategies and how different weather conditions, various control strategies can influence \textit{Culex} population dynamics.
This key feature is essential for public health officials and policymakers aiming to optimise vector control efforts and potentially minimise the spread of mosquito-borne diseases.
The insights from our current study carry important implications for vector management strategies aimed at controlling the \textit{Culex} population at the Cook county,  Illinois, USA. 
Here are some key takeaways that we mention:
Role of temperature and precipitation: our mathematical model underscores the critical role of temperature and precipitation in \textit{Culex} mosquito population dynamics. 
Targeting mosquito control efforts during periods and in areas with temperatures favourable for mosquito activity can potentially optimise interventions.
Adaptability of the weather-driven model: due to its flexibility, our weather-driven mathematical model can be adapted to different regions as new data becomes available. 
This adaptability ensures that control measures remain effective in diverse ecological settings.
Utilising real-time weather data: incorporating real-time weather data can potentially enhance the responsiveness of vector control strategies, thus enabling timely interventions based on current environmental conditions.
Evaluating vector control strategies: Our model can serve as a tool for assessing the effectiveness of various vector control strategies, providing evidence-based recommendations to improve current practices and support the development of new interventions.
By integrating these insights into vector management programs, public health authorities can develop more effective and efficient strategies to reduce mosquito-borne disease transmission and protect public health.

\par
Aligned with previous research efforts, the sensitive parameters identified in our sensitivity tests closely match those reported in the study referenced in \cite{EZANNO201539}. 
For instance, these include mortality rates of adult mosquitoes, transition rates from host-seeking to engorged mosquitoes, and egg-laying rates.
However, there are some exceptions due to differences in the response variables used in their analysis compared to ours.
In our simulations, we perform the sensitivity analysis on $R_{0_{F}}$ and the authors in \cite{EZANNO201539} study the sensitivity associated with peak value, attack rate and 
parity rate.
In general, our results align with the findings presented in \cite{EZANNO201539, CAILLY20127}.
The mathematical form of the basic offspring number aligns with the findings of \cite{YANG_MACORIS_GALVANI_ANDRIGHETTI_WANDERLEY_2009, doi:10.1142/S0218339015500278}, although our research specifically focused on the \textit{Culex} species.
The functional dependence and magnitude of $R_{0_{F}}$ on temperature in our study are very similar to those reported by \cite{YANG_MACORIS_GALVANI_ANDRIGHETTI_WANDERLEY_2009}.
Our simulations also validate the conclusions drawn by the authors in \cite{10.1093/jme/tjad088} that adulticide treatment do not consistently reduce the population of the target WNV vector \textit{Culex}, including the magnitude of $R_{0_{F}}$ as depicted in the figures \ref{fig:abatement1}, \ref{fig:abatement2},  \ref{fig:abatement1R0} and \ref{fig:abatement2R0}.
Our \textit{Culex} trait-based $R_{0_{F}}$ models successfully isolated the physiological effects of temperature on fecundity, and the functional form of $R_{0_{F}}$ aligns qualitatively with the findings of the authors in \cite{10.7554/eLife.58511}. 
Moreover, the graphical representation of $R_{0_{F}}$ versus temperature closely resembles that presented by the same authors.

\par
Given that our model relies on mechanistic ODEs as described in  the sections \ref{Modeldescription}, \ref{ULVSpray1} and accurate data regarding distinct stages in the mosquito life cycle is essential, precise estimation of these parameters can substantially enhance the model's predictive capabilities. 
At present in our work, we approximate the values of various parameters due to insufficient information, but once this data becomes accessible, the accuracy of our model predictions will markedly increase. 
To address this issue, we conduct sensitivity analyses to uncover relationships and quantify the uncertainty inherent in the model's outputs.
However, there were some discrepancies between observed trap data and model predictions. 
This can be due to significant intra-annual variations in the species' abundance. 
Nonetheless, our mechanistic, weather-driven model successfully predicts substantial year-to-year variations in relative mosquito abundance. 
It would be especially important to investigate how the mosquito population in one year affects the population in the following year.
No observed studies on the abundance of overwintering stages of \textit{Culex} species are available to precisely define the beginning and end of the favourable period, thus forcing us to make certain assumptions to run simulations and validate our model. 
Studies specifically aimed at identifying the factors that determine the start of diapause for each epidemiological season can help in more accurately defining the end of the favourable period.
Our model specifically predicts the relative abundance of the blood-feeding stage over time, which not only facilitates the comparison of predictions and observations but also allows for the integration of this population dynamics model with an epidemiological model.
Such coupling can enhance the identification of at-risk periods for pathogen transmission, which depends not only on mosquito abundance but also on pathogen development within the vector and the vector's life cycle, both of which are influenced by weather-driven factors.
Our model does not explicitly include a spatial component, assuming a favourable environment for breeding and maturation into subsequent stages, without accounting for dispersal. However, we do identify and explicitly model the actively moving individuals (emerging, blood-seeking, and egg-laying) along with their emergence rates and periods. 
This approach allows us to potentially incorporate dispersion within our modelling framework.
Even without a spatial component, it is feasible to account for host availability by adjusting the development duration and associated mortality rates. 
This enables us to refine the model without altering its core structure, leaving ample room for further exploration.
In our model, we solely focus on the impact of adulticide, but targeted control measures in other life stages can be assessed using this modelling framework. 
For instance, larvicide can be utilised to decrease the larvae population or disrupt mating, or adjust the duration and schedule of maturation and emergence.
\par
This study offers valuable insights into the influence of weather-driven factors (temperature and rainfall) on the dynamics of \textit{Culex} mosquito populations and the effectiveness of various abatement strategies. 
Our mathematical model illustrates how vector management strategies affect the basic reproduction number ($R_{0_{F}}$) of the \textit{Culex} population within a weather-driven system. However, several avenues for future research could deepen our understanding and improve the practical application of these findings, as outlined below:
Incorporating additional weather-driven factors: While this study focused on temperature and precipitation, other environmental variables, such as habitat type, wind direction, and land use changes, may also significantly impact \textit{Culex} dynamics. 
Future models should integrate these factors to provide a more comprehensive understanding of the ecological drivers of \textit{Culex} population growth.
Expanding the model to other geographic regions: The current model is based on data from Cook County, Illinois. 
Extending it to other areas with different climatic conditions, such as Texas or Florida, would help assess its robustness and adaptability. 
Collaborating with other mosquito abatement districts to gather relevant data could support model validation and highlight regional differences.
Incorporating finer microclimate data: More detailed empirical studies on mosquito survival and feeding behaviour under varying environmental conditions are essential for improving the accuracy of future models.
This information can give us necessary update about the overwintering \textit{Culex} population distribution.  
Including extreme weather events: It is crucial to examine the impact of sudden extreme weather events, such as El Ni\~no, La Ni\~na, or heatwaves, on the \textit{Culex} population. 
Understanding these effects can improve our ability to predict the epidemiological outcomes of mosquito-borne diseases, helping to better inform public health responses and preventive strategies.
Integration with climate change projections: considering future climate scenarios can enable us to get insights into how climate change might affect the distribution and dynamics of \textit{Culex} mosquitoes. 
Inclusion of multiple species: We also acknowledge the subtle differences in species-specific responses to temperature, which will be integrated into future two-species models (\textit{Cx. pipiens} or \textit{Cx. restuans}).
This will aid in anticipating and preparing for future public health challenges related to vector-borne diseases. 
By incorporating climate change projections, future mathematical models can build on our current findings to create more accurate and applicable tools for predicting and managing \textit{Culex} populations.
\par
In conclusion, the weather-driven mathematical model presented in our current work, offers a crucial tool for understanding and managing \textit{Culex} population dynamics. 
By considering meteorological factors and control interventions, our basic mathematical model provides insights that can potentially help to guide vector control efforts and mitigate the impact of mosquito-borne diseases. 
Moving forward, continued research in this research area, coupled with finer field various data while validating the model's predictions, will be an important factor for improving our ability to control \textit{Culex} populations and thus, reduce the burden of associated diseases.

\section{Acknowledgement}
This publication is supported by Cooperative Agreement Number U01CK000651 from the Centers for Disease Control and Prevention and National Institute of Health under Grant No. NIH R01AI125622A. Its contents are solely the responsibility of the authors and do not necessarily represent the official views of the Centers for Disease Control and Prevention and National Institute of Health.

\bibliography{MosPopoModel}

\end{document}